\documentclass[a4paper,11pt]{article}
\usepackage{amsmath,amssymb,amsthm,color}
\usepackage{url}
\usepackage{graphicx}
\usepackage{caption} 

\begin{document}

\title{Quantifying antiproliferative effects of quinolinic acid on melanoma, macrophage and keratinocyte cells using a parametric cell-viability model}
\author{Roumen Anguelov$^1$, Charlise Basson$^2$, Yvette N Hlophe$^2$,\\ Avulundiah Edwin Phiri$^3$, June C Serem$^4$  , Tivoli Visser$^2$, Priyesh Bipath$^2$\\
$^1$Department of Mathematics and Applied Mathematics, University of Pretoria\\
$^2$Department of Physiology, University of Pretoria\\
$^3$Department of Mathematics, Copperbelt University, Kitwe, Zambia\\
$^4$Department of Anatomy, University of Pretoria}
\date{}

\maketitle

\begin{abstract}
This paper presents a robust mathematical framework for quantifying antiproliferative effects from noisy \textit{in vitro} cell viability experiments. The methodology is demonstrated using crystal violet assay measurements of quinolinic acid–induced growth inhibition in B16-F10 murine melanoma, RAW264.7 macrophage, and HaCaT keratinocyte cells through parametric cell viability models. Experimental data exhibited substantial variability and violated the independence assumptions underlying classical inferential statistics. To address these challenges, the proposed framework combines minimal statistical analysis, comprising model-free confidence intervals and pooled within-replicate variability, with deterministic approximation based on least-squares fitting to experimental means and leave-one-replicate-out cross-validation. While all three cell types were described by a common mechanistic framework, each required a distinct parameterisation to capture its characteristic response to quinolinic acid. The resulting one- and two-parameter models accurately described dose- and time-dependent inhibition, with predictive errors close to the intrinsic experimental variability. The models also yielded explicit expressions for time-dependent $IC_{50}$ values, enabling reliable prediction of inhibitor concentrations required to achieve specified levels of growth inhibition. The proposed framework provides a practical and robust approach for analysing noisy preclinical cell viability data and can be readily extended to other antiproliferative agents and experimental systems.
\end{abstract}


\section{Introduction}

The inhibition of tumor and cancer cell growth is a central objective in preclinical drug development. Quantitative evaluation of the antiproliferative effects of candidate inhibitors is typically achieved through \textit{in vitro} experiments, in which cell viability is measured over time across a range of inhibitor concentrations to assess drug efficacy and characterize dose-dependent responses.
Such experiments, however, are inherently subject to considerable random variability that arises from biological fluctuations, technical limitations (e.g., pipetting errors, inadequate resuspension leading to cell clumping), and uncontrolled environmental factors. Consequently, the resulting data often appear “noisy”: trends that are visible in one experimental replicate may not be fully reproducible in another, and measurements may exhibit non-systematic deviations that do not reflect the true underlying dose–time response of the cell population.

A thorough statistical analysis of these data is challenging for two main reasons. First, classical inferential methods (e.g., \(t\)-tests, \(F\)-tests, linear regression) typically assume normality of the errors and independence of all observations. In the experimental setting considered here, the measurements corresponding to different inhibitor concentrations within the same replicate are normalized by a common control population, which introduces dependence across concentration, while the measurements at different times are independent. Moreover, the number of independent replicates is often small (e.g., \(K=4\)), making it difficult to verify distributional assumptions reliably. Second, even if one attempts to build confidence intervals or hypothesis tests, the correlation structure and the non-linearity of the expected response curve complicate the application of standard parametric procedures.

In this paper we present a methodology that combines a minimal set of statistical tools with a deterministic, approximation-based approach to derive and validate mathematical models of cell growth inhibition. The statistical component is limited to the construction of model-free confidence intervals for the mean viability at each concentration–time combination, and to the calculation of a pooled within-replicate standard deviation. All further steps treat the problem as a deterministic approximation problem: the quantity of interest is the unknown mean viability function \(\mu(c,t)\), and we aim to find a suitable parametric model \(\phi(a;c,t)\) that approximates \(\mu(c,t)\) with high fidelity, without relying on probabilistic assumptions about the residuals. Parameter estimation is performed by least-squares fitting to the experimental means (averaged over replicates), and model validation is carried out using leave-one-replicate-out cross-validation, comparing the resulting prediction error to the inherent experimental scatter.

The approach is exemplified on the modelling and analysis of the antiproliferative effects of quinolinic acid on three distinct cell populations: B16-F10 murine melanoma, RAW264.7 macrophages, and human HaCaT keratinocytes. Quinolinic acid (Quin) is a metabolite of the kynurenine pathway, primarily derived from tryptophan catabolism. Its biological activity is closely linked to cellular energy metabolism and redox balance. In the context of cancer and immune-related cells, Quin exerts antiproliferative effects through several interconnected mechanisms such as oxidative stress, mitochondrial dysfunction and altered NAD$^+$ pathways, \cite{Perez2012,Yan2024}. Although the same general modelling framework is applied to each cell line, the estimated parameters differ substantially, reflecting the different sensitivity of each cell type to the inhibitor. The analysis shows that, despite the high variability of the raw data, the derived models capture the essential trends and provide reliable predictions of the mean viability. Interestingly, the three cell lines require differently parameterised model structures, illustrating the flexibility of the proposed methodology.

The remainder of this paper is organised as follows: Section~2 describes the experimental setup, the data collection and initial statistical analysis. Section~3 presents the derivation of general mathematical model of cell viability, the method of fitting it to experimental data as well as means of validating the model. Section~4 details the specific models for each one of the considered cell populations, including validation of the models. Section~5 contains discussion on the significance and the practical application of the models. Section~6 concludes the paper with remarks on the strengths and limitations of the approach.

\section{Experimental setting and data analysis}

\subsection{Crystal violet assay}\label{Subsection_CV}
Cell viability is commonly used as a measurement in \textit{in vitro} testing of anti-cancer effects of treatments, e.g. impact of an inhibitor on the cancer cell growth. Cell viability is defined as the percentage ratio of viable cells in a treated (inhibited) population to the viable cells in an untreated control population at a given time point.

The cell viability under inhibition at different concentrations of the inhibitor (Quin) and at different duration of exposure is measured using the crystal violet assay,  \cite{CVA2016}. Specifically, melanoma B16F10, RAW264.7 macrophages and keratinocyte (HaCaT) cells were seeded in well plates at seeding densities of 5000, 10000 and 10000 cells per well, respectively. The plates were incubated overnight at 37°C to allow the cells to adhere to the wells. Each experimental run involves 54 wells, divided in 6 groups of 9 each. After the incubation period, the cells in the five of the groups are treated with the five concentrations of Quin indicated in section 2.5. One group is not treated with Quin and is used for control. At the specified times, namely after 24, 48 and 72 hours of treatment, three wells from each group were processed as follows: 200 µL of 1\% glutaraldehyde is added to each well. The plates are incubated at 37°C for 30 minutes, after which the glutaraldehyde is removed. Subsequently, 100 µL of 0.1\% crystal violet was added, and the cells are stained at room temperature for 30 minutes. Then the plates were rinsed with water and left to air dry completely. Once dry, the bound crystal violet is solubilised using 100 µL of 10\% acetic acid. Absorbance was measured at 630 nm using an EPOCH microplate reader.

The size of the cell population strongly and positively correlates with dye absorbance. A measurement of the cell viability is calculated as the absorbance of the treated population divided by the absorbance of the control population. Each of the two absorbance numbers is the average of the readings of the respective three wells.

\subsection{Statistical approach to data analysis}\label{Subsection_Data}

The measurements of cell viability exhibit substantial variability due to the nature of the experiment. Hence, for a more reliable estimation, the experiment is repeated several times for the same concentrations and duration of exposure. Let the experiments be repeated $K$ times, for specified concentrations $c_i$, $i=1,...,I$, and durations of exposure $t_j$, $j=1,...,J$. In the specific setting described in Subsection \ref{Subsection_CV}, we have $I=5$, $J=3$, $K=4$, but the discussion below applies to any integer values of $I$, $J$ and $K\geq 3$. The measurement $v_{ijk}$ of the cell viability can be considered as observation of a random variable $V_{ijk}$. So, we have the data set
 \begin{equation}\label{dataSet}
 \{v_{ijk}:k=1,...,K,\ j=1,...,J,\ i=1,...,I\}
 \end{equation}
 In statistical terms we consider the cell viability for concentration $c$ and time $t$ (as measured by the crystal violet assay) to be a random variable with a mean $\mu(c,t)$. Then $E(V_{ijk})=\mu(c_i,t_j)$. Further, we make the following assumptions
 \begin{itemize}
 \item[(a)] For a fixed $c_i$, the random variables $\{V_{ijk}:k=1,...,K,\ i=1,...,I\}$ are all independent and have the same variance (homoscedasticity).  This follows from the experimental design because each replicate is an independent execution of the entire experiment (including a fresh control), measurements for different concentrations within the same replicate are correlated, but for a fixed concentration the replicates across different experiments are independent. Further, to confirm the homoscedasticity, we conduct
Levene’s test, where we obtain $p$ values well above 0.005. We note that the independence for different $c_i$ is problematic because they use comparison with the same control population ($c=0$).
 \item[(b)] The random variables $V_{ijk}$ are normally distributed. It is an expected property, but it also confirmed by a Shapiro-Wilk normality test of the data.
  \end{itemize}

\subsection{Confidence intervals}\label{Subsection_CI}
Under the conditions (a) and (b) in Subsection \ref{Subsection_Data}, we construct $\alpha$ confidence interval for $\mu(c_i,t_j)$ as follows. Let us fix $c_i$.
From the independence and normality of $\{V_{ijk}:k=1,...,K,\ j=1,...,J\}$ it follows that
\[
\frac{\sum\limits_{j=1}^J\sum\limits_{k=1}^{K}(V_{ijk}-\bar{V}_{ij\bullet})^2}{\sigma_i^2}
\]
is a $\chi^2$ random variable with $J(K-1)$ degrees of freedom. Therefore, for every $j=1,...,J$ the random variable
\[
T_{ij}=\frac{\bar{V}_{ij\bullet}-\mu(c_i,t_j)}{\sqrt{\frac{\sum\limits_{j=1}^J\sum\limits_{k=1}^{K}(V_{ijk}-\bar{V}_{ij\bullet})^2}{JK(K-1)}}}=
\frac{\frac{\bar{V}_{ij\bullet}-\mu(c_i,t_j)}{\sigma_i/\sqrt{K}}}{\sqrt{\frac{\sum\limits_{j=1}^J\sum\limits_{k=1}^{K}(V_{ijk}-\bar{V}_{ij\bullet})^2}{\sigma_i^2J(K-1)}}}
\]
has a Student $t$ distribution with $J(K-1)$ degrees of freedom. The quantity
\[
s^2_{p,i}=\frac{\sum\limits_{j=1}^J\sum\limits_{k=1}^{K}(v_{ijk}-\bar{v}_{ij\bullet})^2}{J(K-1)}
\]
is called pooled sample variance, and $s_{p,i}$ - pooled sample standard deviation, for the data at $(c_i,t_1)$,...,$(c_i,t_J)$.
 Let $[-\nu,\nu]$ be an $\alpha$-confidence interval for a $t$ distributed random variable with $J(K-1)$ degrees of freedom. Then
\begin{equation}\label{eq_CI}
\left[\bar{v}_{ij\bullet}-\nu\frac{s_{p,i}}{\sqrt{K}},\bar{v}_{ij\bullet}+\nu\frac{s_{p,i}}{\sqrt{K}}\right]
\end{equation}
is an $\alpha$-confidence interval for $\mu(c_i,t_j)$. Note that all these confidence intervals are not related to any model and all have the same width of $2\nu\frac{s_{p,i}}{\sqrt{K}}$ for every $i$. Let us note that confidence intervals can be constructed for any $c_i$ and $t_j$ by using only the replicates for these values of concentration and time. The advantage of combining the data for $t_1,...,t_J$ in the stated way is that we obtain a Student $t$ random variable with larger number of degrees of freedom, which in turn reduces the average with of the confidence intervals.

\begin{figure}[ht!]
\includegraphics[scale=0.3]{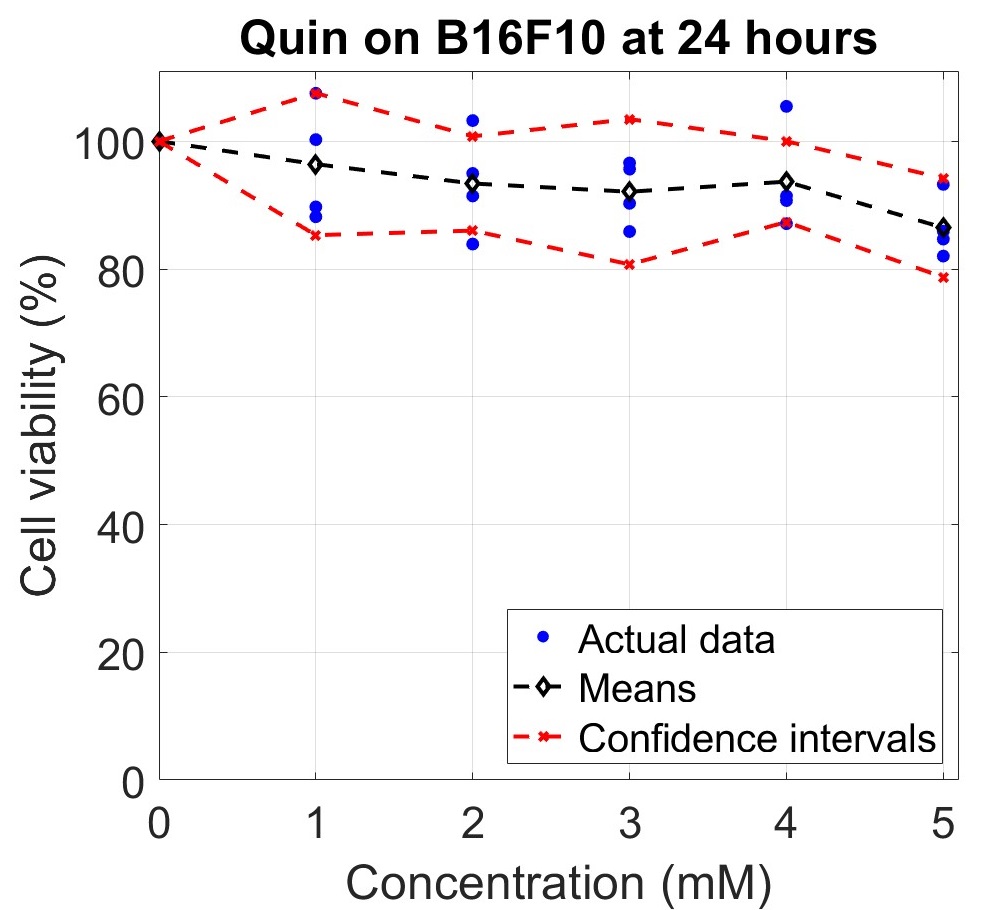} \includegraphics[scale=0.3]{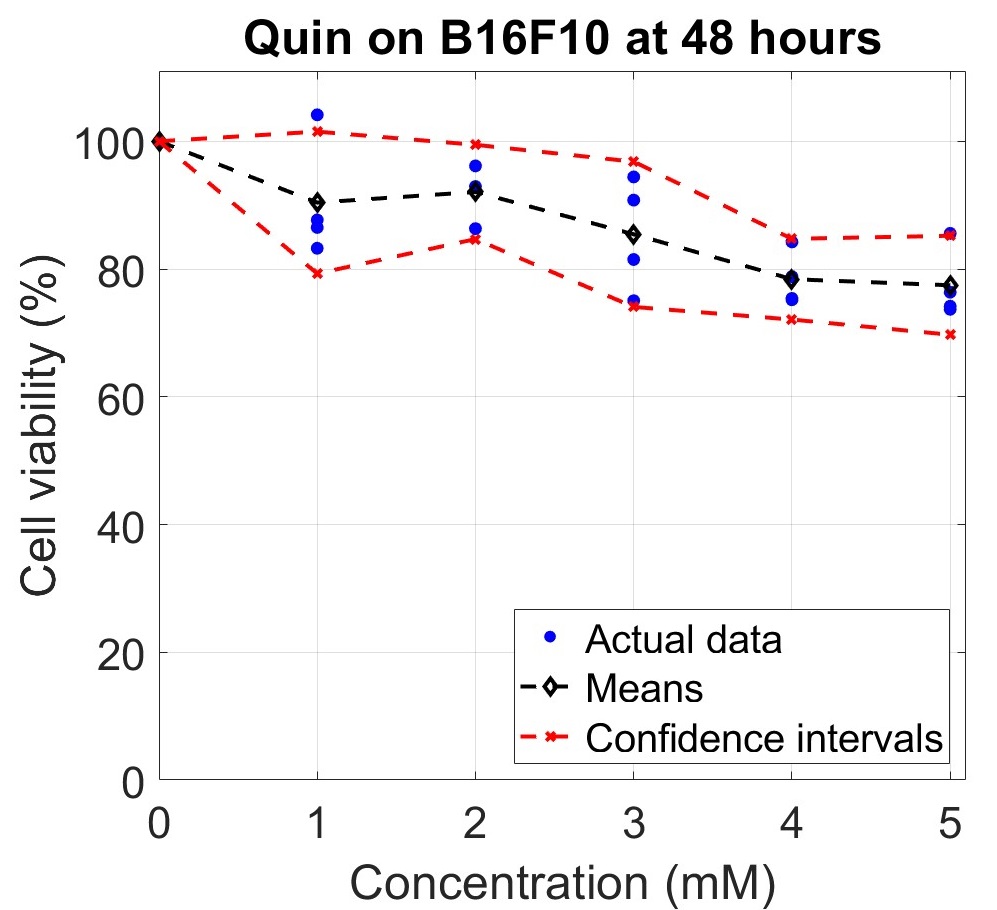}
\includegraphics[scale=0.3]{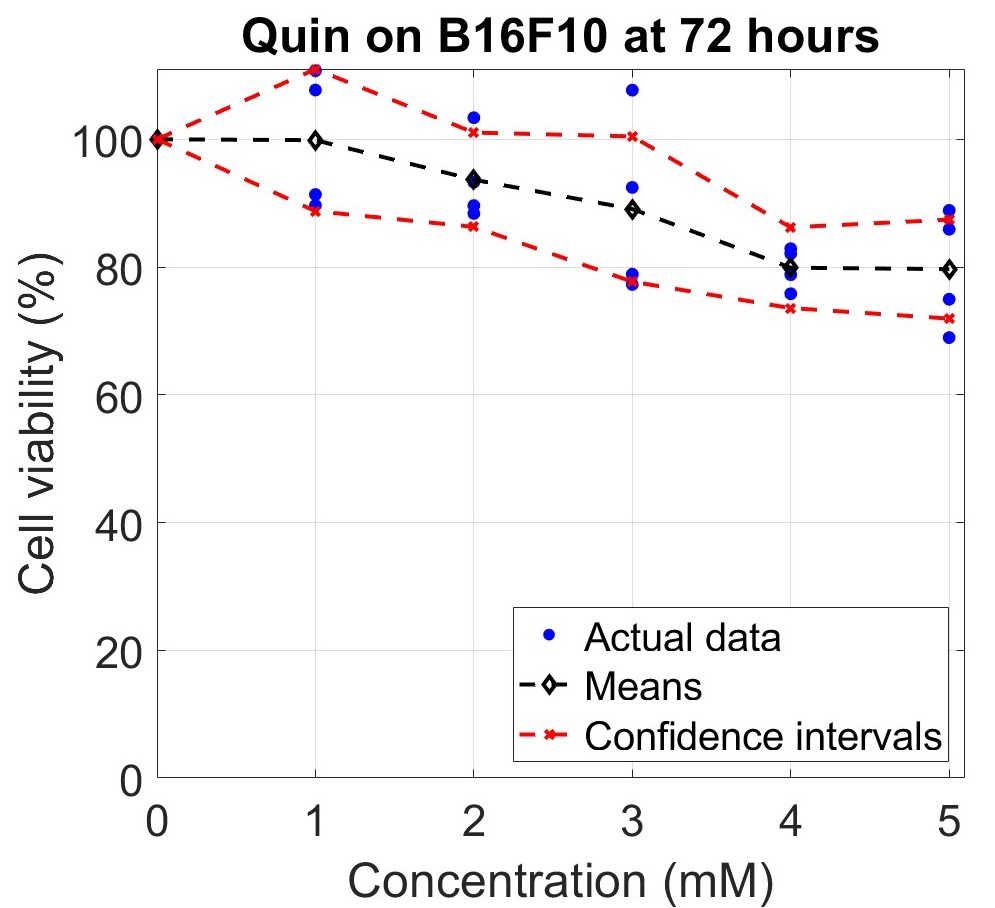}
\caption{Experimental measurements, means and confidence intervals of the cell viability of B16F10 cell populations under inhibition of Quin at different concentrations}\label{Data_B16F10}\vspace{5mm}

\centering\includegraphics[scale=0.315]{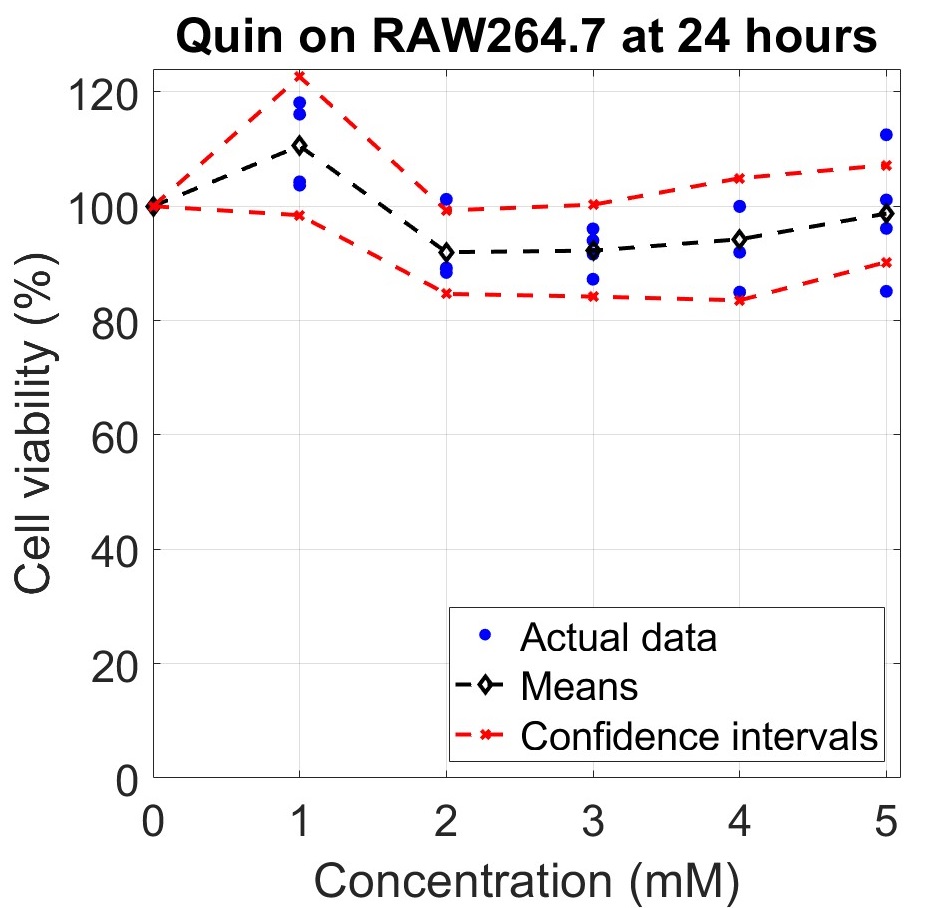} \includegraphics[scale=0.315]{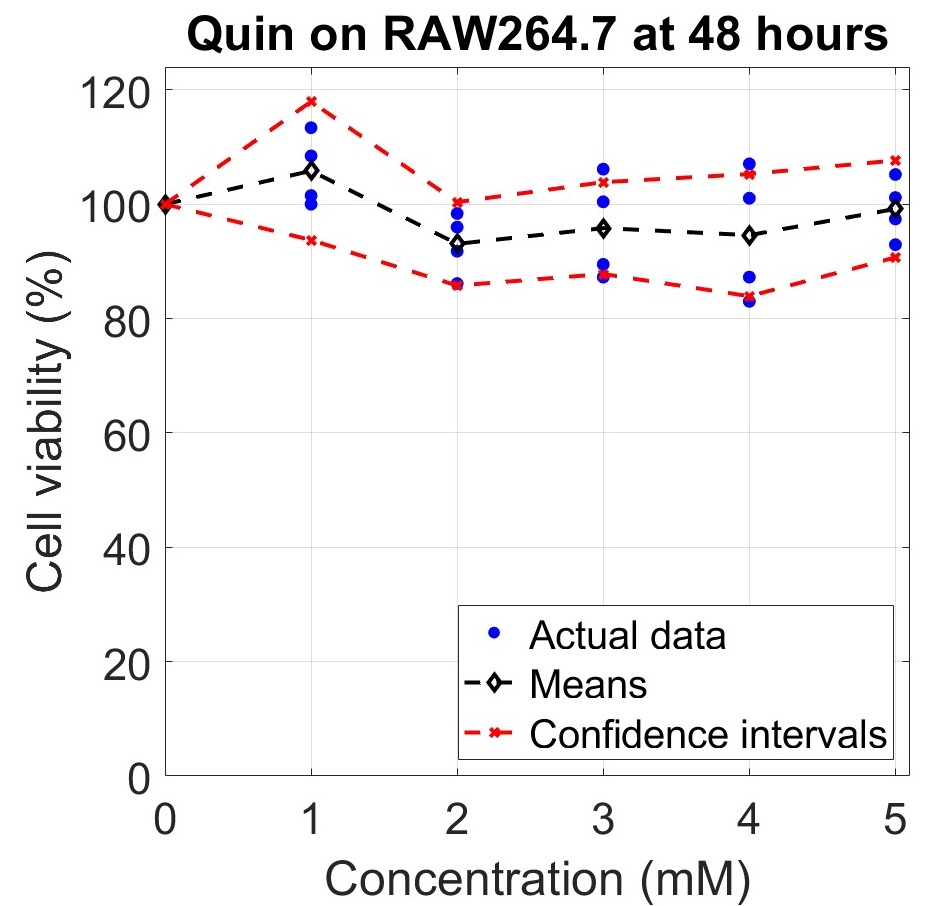}
\includegraphics[scale=0.315]{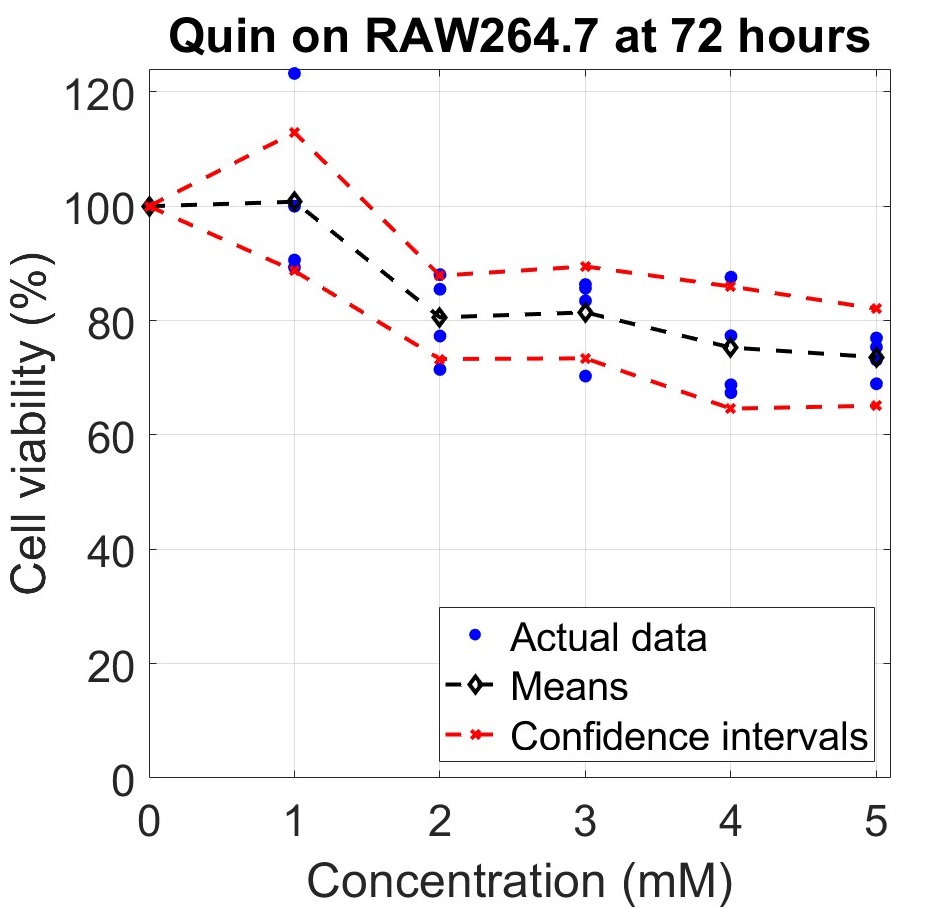}
\caption{Experimental measurements, means and confidence intervals of the cell viability of RAW264.7 cell populations under inhibition of Quin at different concentrations}\label{Data_RAW}\vspace{5mm}

\centering\includegraphics[scale=0.315]{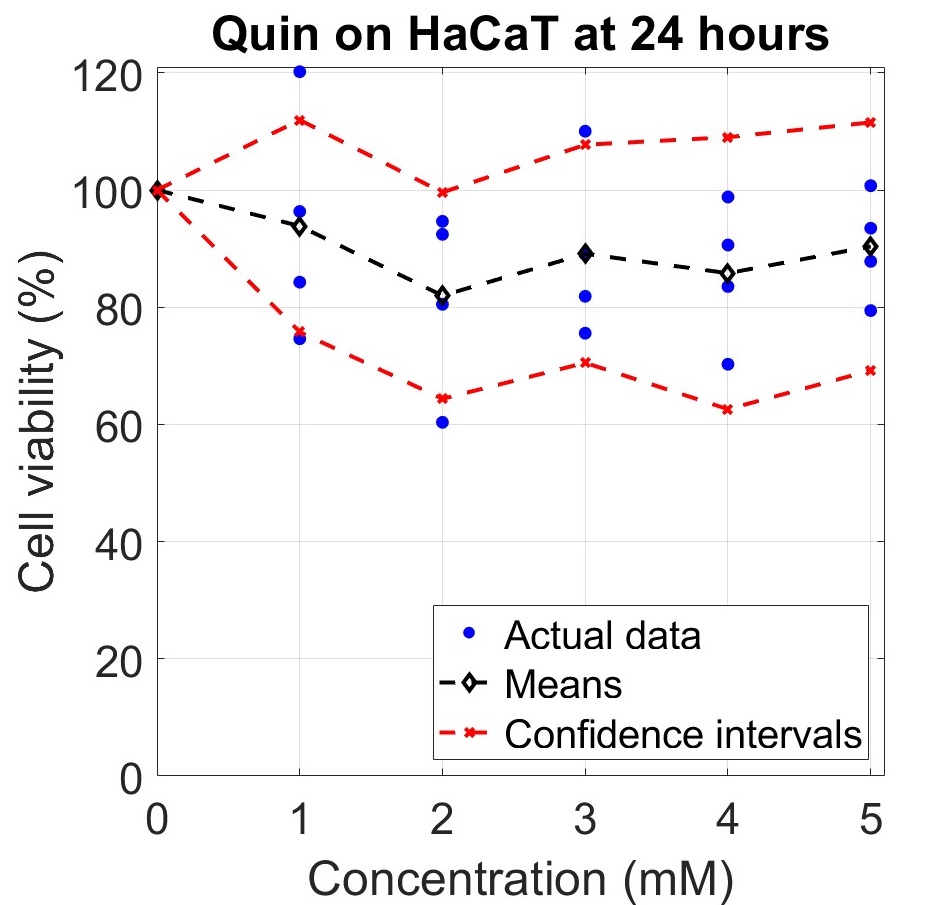} \includegraphics[scale=0.315]{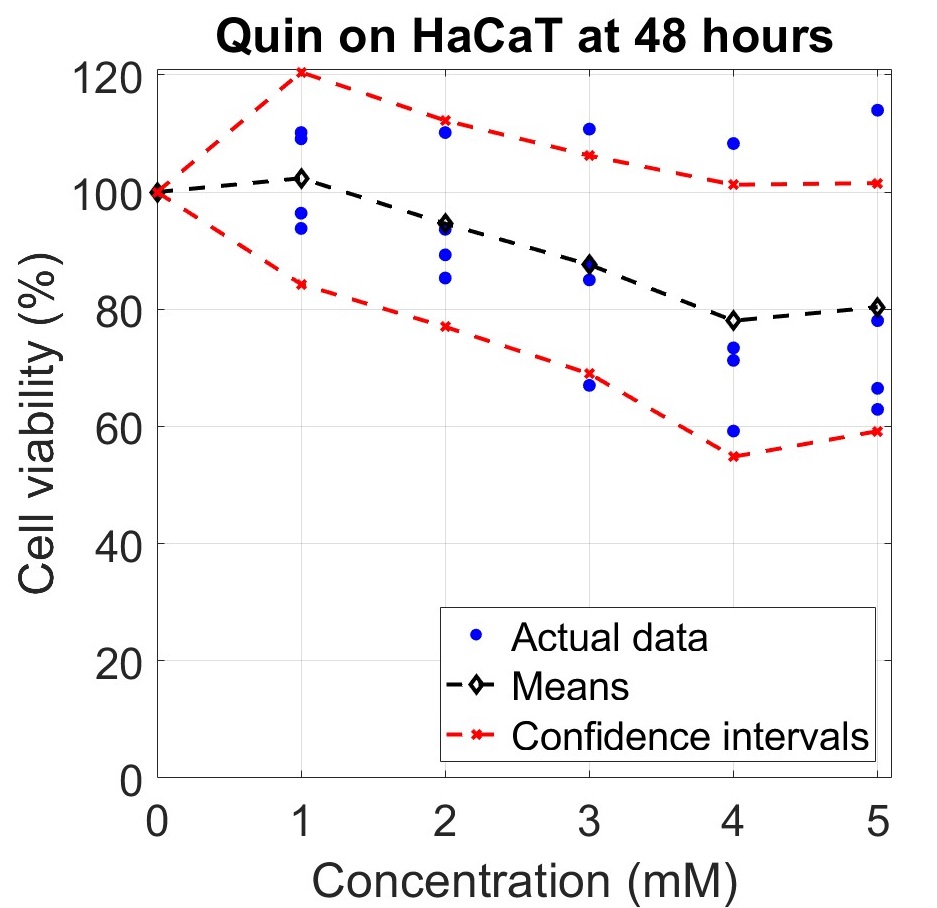}
\includegraphics[scale=0.315]{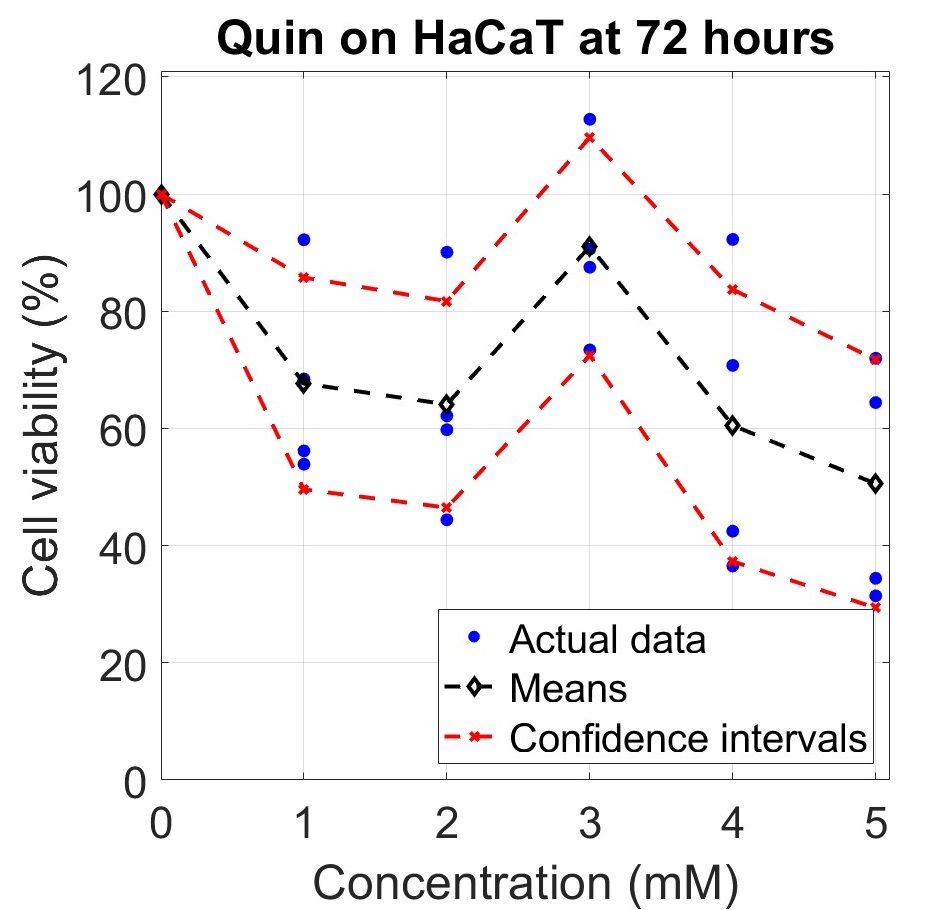}
\caption{Experimental measurements, means and confidence intervals of the cell viability of HaCaT cell populations under inhibition of Quin at different concentrations}\label{Data_HaCat}
\end{figure}

Figure \ref{Data_B16F10}, Figure \ref{Data_RAW} and \ref{Data_HaCat} represent the data, means and 95\% confidence intervals of the cell viability of B16F10, RAW264.7 and HaCaT cell populations under inhibition by Quin. One can observe the significant variability of data, e.g., the range of data at some times and concentrations is wider than the range of the means over all times and concentrations. This is an important motivation for the careful and rigorous statistical and deterministic analysis with the goal to extract biologically meaningful trend and to reliably quantify of the effect of inhibition. An initial step in this process is constructing 95\% confidence intervals using the pooled variance as described in \eqref{eq_CI}. It can be observed that these confidence intervals exclude some data points and that a general trend of decrease with respect to $c$ and $t$ is more apparent, e.g., the confidence intervals for the larger concentrations and duration of exposure do not contain 100\%. The latter fact is by itself an indicator of inhibition effect. Let us remark that we cannot use pooled variance over all data to further reduce the with of the confidence intervals as the measurements are not independent across concentrations. In the next Subsection we discuss a more reliable measure of  inhibition.

\subsection{Significance of inhibitory impact}\label{Subsection_Impact}

The goal is to establish if the experimental data provides evidence of inhibitory impact on the tested cell population.
Let $c_i$ be fixed. Consider the null hypothesis
\begin{equation}\label{HypTest}
H_0: \mu(c_i,t_j)=100,\text{ for all } j=1,...,J,
\end{equation}
versus the alternative that $H_0$ is not true. Under $H_0$, each of the random variables
\[\frac{V_{ij\bullet}-100}{\sigma_i/\sqrt{K}},\ j=1,...,J,\]
has standard normal distribution. Then, using their independence, we have that
\[
\frac{\sum\limits_{j=1}^J(V_{ij\bullet}-100)^2}{\sigma_i^2/K},
\]
is $\chi^2$ random variable with $J$ degrees of freedom. Hence,
\[
F=(K-1)K\frac{\sum\limits_{j=1}^J(V_{ij\bullet}-100)^2}{\sum\limits_{j=1}^J\sum\limits_{k=1}^{K}(V_{ijk}-\bar{V}_{ij\bullet})^2}
=\frac{\frac{\sum\limits_{j=1}^J(V_{ij\bullet}-100)^2}{J\sigma_i^2/K}}{\frac{\sum\limits_{j=1}^J\sum\limits_{k=1}^{K}(V_{ijk}-\bar{V}_{ij\bullet})^2}{J(K-1)\sigma_i^2} },
\]
is $F$ random variable with $J$ and $J(K-1)$ degrees of freedom. This random variable is the test statistics for the test \eqref{HypTest}. The observed value is calculated from the data and the $p$-value of the test is obtained from the respective $F$ distribution. This is typically a first test to be conducted, as there is no reason to model inhibition if there is no sufficient evidence that it exists, e.g. $p<0.05$.

The test \eqref{HypTest} is conducted for each cell population, B16F10, RAW264.7, HaCaT, at the concentrations used in the experimental data, namely $1mM$, $2mM$, $3mM$, $4mM$, $5mM$. The calculated $p$-values are presented in Table \ref{tableHypTest}. At 5\% level of significance the null hypothesis is rejected for 4 out of the 5 concentrations for each cell population. This result motivates the construction of a mathematical model to quantify the inhibitory impact conducted in the next section.

\begin{table}[h]
\caption{The $p$-values for the hypothesis test \eqref{HypTest} at the individual concentrations of Quin for each of the three cell populations. The null hypothesis in \eqref{HypTest} is rejected at level of significance of 5\% for  4 out of the 5 concentrations for each of the three cell populations. The respective $p$-values are given in bold.}\label{tableHypTest}\vspace{3mm}

\centering\begin{tabular}{|c|c|c|c|c|c|}\hline &&&&&\\[-8pt]
concentration&1mM&2mM&3mM&4mM&5mM\\[-8pt]&&&&&\\\hline &&&&&\\[-8pt]
B16F10 &0.29171 &\textbf{0.03245} &\textbf{0.02346}  &\textbf{0.00002} &\textbf{0.00004} \\[-8pt]&&&&&\\\hline &&&&&\\[-8pt]
RAW264.7&0.23660&\textbf{0.00064}&\textbf{0.00217}&\textbf{0.00309}&\textbf{0.00052}\\[-8pt]&&&&&\\\hline &&&&&\\[-8pt]
HaCaT&\textbf{0.01841}&\textbf{0.00445}&0.22961&\textbf{0.00952}& \textbf{0.00220}   \\[-8pt]&&&&&\\\hline
\end{tabular}
\end{table}

\section{Mathematical modeling of inhibition}

\subsection{A generic model}\label{FirstSubsection}

The inhibition of tumor growth is typically modeled as a reduction of its natural growth rate, \cite{Jarret2018}, \cite{Siv}, \cite{Yin2019}. Using the so called "log-kill" pattern, \cite{Skipper1964}, \cite[Eq.(7)]{Siv}, \cite[Eq.21, Eq.22]{Yin2019}, the growth rate of the inhibited population is given in the form $r-h$, where $r$ is the natural growth rate under optimal conditions and $h$ represents the growth rate reduction due to the impact of the inhibitor. The rate reduction $h$ may vary with time $t$ and the concentration $c$ of the inhibitor, \cite{Siv,Panetta1997,mmnp2025,article1}.

In general, the function $h$ results from the impact of the inhibitor to complex signalling network and metabolic processes. While the possible mechanisms of this impact are at least partially known, quantification of the inhibitor induced causal relations in this complex setting is seldom available. Combining biological knowledge with insight from the data and reliable model fitting produced models, which were useful in similar seedings, \cite{Amitans2025,mmnp2025,article1,article2}. In this paper we follow the same strategy to (i) determine a general form of $h$ as a function of $c$ and $t$ depending on a parameter vector $a$, and (ii) calculate an optimal, in a specified sense, value of $a$.
Considering that $h$ represents the impact of inhibition, we have
\begin{itemize}
\item $h(0,t)=h(c,0)=0$, $t\geq 0$, $c\geq 0$.
\item $h$ is monotone increasing in both $c$ and $t$;
\item there exists $m(c)>0$ such that $\displaystyle\lim_{t\to+\infty}h(c,t)=m(c)$.
\end{itemize}

\subsection{Mathematical model of the cell viability}\label{Subsection_Model}

We recall that the cell viability is defined as the percentage ratio of viable cells in a treated (inhibited) population to the viable cells in an untreated control population at a given time point.

Let us consider first a model for the growth of the control cell populations, that is, when the cells are provided with optimal growth conditions not subjected to inhibition. Under these conditions, the population can be modelled via constant growth rate model, \cite[Table 1]{Jarret2018}, \cite[Table 1, Eq. 20]{Yin2019}, \cite[Eq.(1)]{Siv},
\begin{equation}\label{eqP}
\frac{dP}{dt}=rP,
\end{equation}
where $P(t)$ is the size of the population at time $t$ and $r$ is the constant relative growth rate per unit time. The solution of \eqref{eqP} is $P(t)=P(0)e^{rt}$. Hence, \eqref{eqP} is also called exponential growth model. This model is applicable while the stated optimal growth conditions prevail.

The size of the inhibited population $Q$ depends on the concentration $c$ of the inhibitor and the time $t$. It satisfies an equation similar to \eqref{eqP}, but with a reduced growth rate. More precisely, we have
\begin{equation}\label{eqQ}
\frac{\partial Q(c,t)}{\partial t}=(r-h(c,t))Q(c,t).
\end{equation}
Taking into account that the initial population of treated and control wells are the same, the solution of \eqref{eqQ} is
\begin{equation}\label{solQ}
Q(c,t)=P(0)e^{\displaystyle rt-\int_0^th(c,\theta)d\theta}.
\end{equation}
Then, from the definition of cell viability it follows that at time $t$ and under inhibitor concentration $c$, it is given by the function
\begin{equation}\label{eqCelVia}
\Psi(c,t)=100 \frac{Q(c,t)}{P(t)}=100e^{-\displaystyle\int_0^t h(c,\theta)d\theta}.
\end{equation}

Let us recall that in Sections \ref{Subsection_Data}--\ref{Subsection_Impact} we considered the cell viability as a random variable. From statistical point of view $\Psi(c,t)$ is precisely the mean of this random variable, namely, $\mu(c,t)$.

\subsection{Cell viability as a function of time}

Considering that we have the cell viability function in the form \eqref{eqCelVia}, the mathematical model requires constructing the function $h$ and estimating the values of parameters that are involved. An appropriate function identified in modeling of inhibition, and satisfying the bullets in Subsection \ref{FirstSubsection}, is
\begin{equation}\label{eq_h}
h(c,t)=m(c)(1-e^{-\lambda(c)t}),
\end{equation}
which is often used to represent saturation in time. The quantity $m(c)$ is the maximum inhibition at concentration $c$, while $\lambda(c)$ describes how fast the inhibition gets close to this maximum. Equation \eqref{eq_h}  yields
\begin{equation}\label{eqCelVia2}
\Psi(c,t)=100 e^{-m(c)\displaystyle\frac{e^{-\lambda(c)t}-1+\lambda(c)t}{\lambda(c)}}.
\end{equation}
In practice, as specific cases, $h$ can be considered as independent of $t$ or as linear on $t$. More precisely,
\begin{itemize}
\item[(i)] if $\lambda(c)$ is very small we have $h(c,t)\approx\lambda(c)t$ and, respectively,
\begin{equation}\label{eqCelVia3}
\Psi(c,t)\approx 100 e^{-\frac{1}{2}m(c)\lambda(c)t^2}=100 e^{-\tilde{m}(c)t^2},
\end{equation}
where $\tilde{m}(c)=\frac{1}{2}m(c)\lambda(c)$. The motivation is only numerical and it has been discussed in \cite{article1}.
\item[(ii)] if $\lambda(c)$ is very large, than we have $h(c,t)\approx 1$ and, respectively,
\begin{equation}\label{eqCelVia4}
\Psi(c,t)\approx 100 e^{-m(c)t},
\end{equation}
depends (as in (i)) on a single function of $c$ to be identified from data. The biological interpretation of this form of the cell viability function is that for large $\lambda$ the maximum inhibition is reached in such a short time that it can be neglected.
\end{itemize}

The form of $m$ and $\lambda$ as functions of $c$ depends on the mechanism the tested inhibitor interacts with the cell population and varies significantly. In similar settings, as presented in \cite{Amitans2025,mmnp2025,article1,article2}, the functions $m$ and $c$ are determined from insight about the inhibition process and the data itself. For the general analysis in this section, let us assume that $m$ and $\lambda$ are given explicitly as functions of $c$ in terms of a vector $a$ of parameters. Hence, the cell viability function is sought in the form
\begin{equation}\label{eqCelVia5}
\Psi(c,t)=\varphi(a;c,t).
\end{equation}

\subsection{The least squares fitting of nonlinear models to data of repeated experiments}

We consider the fitting of cell viability function $\Psi(c,t)$ in its parameterised form $\varphi(a;c,t)$ to experimental data using the Least Squares Method (LSM).
The objective function is the sum of the squared errors (SSE)
\begin{eqnarray}
\text{SSE}(a)&=& \sum_{i=1}^I\sum_{j=1}^J\sum_{k=1}^{K}(\varphi(a;c_i,t_j)-v_{ijk})^2\nonumber\\
&=&K\sum_{i=1}^I\sum_{j=1}^J(\varphi(a;c_i,t_j)-\bar{v}_{ij\bullet})^2+\sum_{i=1}^I\sum_{j=1}^J\sum_{k=1}^{K}(v_{ijk}-\bar{v}_{ij\bullet})^2.
\label{LSM_objective}
\end{eqnarray}

Clearly, minimizing the LMS objective function is equivalent to minimizing
\begin{equation}\label{LSM_objective_means}
\text{SSE}_{\text{means}}(a)=\sum_{i=1}^I\sum_{j=1}^J(\varphi(a;c_i,t_j)-\bar{v}_{ij\bullet})^2,
\end{equation}
since the second term of in \eqref{LSM_objective} does not depend on $a$ or $\varphi$. This implies that the optimal value for $a$
\begin{equation}\label{opt_a}
\hat{a}=\arg\min \text{SSE}_{\text{means}}(a)=\arg\min\text{SSE}(a)
\end{equation}
is determined only by the mean values of all experimental measurements for any $(c_i,t_j)$.
As mentioned at the end of Section \ref{Subsection_Model},  $\Psi(c,t)=\mu(c,t)$. From this point of view, the result \eqref{opt_a} can be expected, since the means $\{\bar{v}_{ij\bullet}:i=1,...,I,\ j=1,...,J\}$ are estimates of $\mu(c_i,t_j)$. Further, the second term in \eqref{LSM_objective} depends on only on the sample variances of replicates given by
\[
s^2_{ij}=\frac{1}{K-1}\sum_{k=1}^{K}(v_{ijk}-\bar{v}_{ij\bullet})^2,
\]
which, under the assumption of normality are independent from the respective means.
Hence, we can re-state LSM as finding $\hat{a}$ minimizing $\text{SSE}_{\text{means}}(a)$ in \eqref{LSM_objective_means}, where the error is given by the root mean square error of the means, namely,
\[
\text{RMSE}_{\text{means}}=\sqrt{\frac{\text{SSE}_{\text{means}}(\hat{a})}{IJ}}.
\]
We note that $\text{RMSE}_{\text{means}}$ represents the average, in the sense of quadratic average, deviation of the means $\bar{v}_{ij\bullet}$ from the function $\varphi(\hat{a};c,t)$ and is an appropriate measure of the accuracy of the model. For the original LSM objective function the error would be
\[
\text{SSE}(a)= \text{SSE}_{\text{all}}(a)= K\times\text{SSE}_{\text{means}}(a)+(K-1)IJ\times s^2_{\text{within}},
\]
where
\[
s_{\text{within}}=\sqrt{\frac{1}{IJ}\sum_{i=1}^I\sum_{j=1}^Js^2_{ij}}
\]
is the average sample standard deviation within replicates.
Hence,
\[
\text{RMSE}_{\text{all}}= \sqrt{\frac{\text{SSE}_{\text{all}}(\hat{a})}{IJK}}=\sqrt{\text{RMSE}_{\text{means}}^2+\frac{K-1}{K}s^2_{\text{within}}}
\]
could be a significant and misleading overestimation of the actual accuracy of the model. More precisely, $\text{RMSE}_{\text{all}}$ includes the experimental noise and thus does not reflect the model’s ability to approximate the mean. RMSE$_{\text{means}}$ is the appropriate measure for the deterministic approximation.

The LSM model fitting in this paper is implemented in MATLAB using the fmincon function.

\subsection{Model validation}\label{sec_validation}

To evaluate the predictive reliability of the model without relying on statistical distribution assumptions, we perform leave-one-replicate-out cross-validation (LOOCV) as follows.

For each replicate \(k = 1,\dots,K\):
\begin{itemize}
    \item[(i)] Remove all data belonging to replicate \(k\) and from the remaining \(K-1\) replicates, compute the cell means
    \(\bar{v}_{ij\bullet}^{(-k)}\).
    \item[(ii)] Fit the model to these means, yielding parameter estimate \(\hat{a}^{(-k)}\).
    \item[(iii)] For every measurement in the left-out replicate, compute the prediction error
    \(v_{ijk} - \phi(\hat{a}^{(-k)};c_i,t_j)\).
\end{itemize}
The overall prediction error is summarized by the root mean squared prediction error
\begin{equation}\label{RMSPE}
\text{RMSPE} = \sqrt{\frac{1}{K  I  J}\sum_{k=1}^{K}\sum_{i=1}^{I}\sum_{j=1}^{J}
\bigl( v_{ijk} - \varphi(\hat{a}^{(-k)};c_i,t_j) \bigr)^2 }.
\end{equation}

A natural benchmark for the irreducible experimental variability is the average within-replicate standard deviation
$s_{\text{within}}$. It measures the typical scatter of individual measurements around their own mean and is independent of any model.
If a model were perfect (\(\varphi(\hat{a}) = \mu\)), then predicting a new replicate would incur an error equal to \(s_{\text{within}}\).
Therefore, the LOOCV prediction error in \eqref{RMSPE} is compared to \(s_{\text{within}}\).
More precisely,
\begin{itemize}
    \item \(\text{RMSPE} \approx s_{\text{within}}\): the model predicts new replicates as well as the experimental repeatability allows – excellent.
    \item \(\text{RMSPE}_{\text{LOOCV}}\) moderately larger than \(s_{\text{within}}\): acceptable, but some systematic lack of fit remains.
    \item \(\text{RMSPE}_{\text{LOOCV}} \gg s_{\text{within}}\): the model fails to capture the underlying trend.
\end{itemize}

The quantity $s_{within}$ from statistical view point is the pooled standard deviation across times and concentrations.
The shared control at a given time violates independence across concentrations needed for statistical inference. Hence, it invalidates any application to inferential statistics, e.g., $t$-tests, $F$-tests.

However, $s_{within}$ is considered from a deterministic perspective, namely, it measures the average spread of individual measurements around their cell-specific mean. No independence assumption is needed to compute it. Thus, using $s_{within}$ as a benchmark for LOOCV is valid, regardless of the correlation structure across concentrations. Hence, the ratio $\frac{\text{RMSPE}}{s_{\text{within}}}$ is a meaningful deterministic measure of predictive accuracy relative to the inherent experimental scatter.

Importantly, and as mentioned, $s_{\text{within}}$ is computed from the data only, independently from any model. The values of $s_{\text{within}}$ for the three cell population is given in Table \ref{swithin}.

\begin{table}[h]
\caption{The values of $s_{\text{within}}$ for the experimental data on B16F10, RAW264.7 and HaCaT.}\label{swithin}\vspace{3mm}

\centering\begin{tabular}{|c|c|c|c|}\hline &&&\\[-8pt]
&B16F10&RAW264.7&HaCaT\\[-8pt]&&&\\\hline &&&\\[-8pt]
$s_{\text{within}}$ &7.9790&8.3900&17.550\\[-8pt]&&&\\\hline
\end{tabular}
\end{table}

\section{Deriving specific inhibition models for B16F10, RAW264.7 and HaCaT cell populations}

\subsection{Data-motivated parametrization of the models.}

We start with fitting the general model \eqref{eqCelVia2} with only assumption of monotonicity of the functions $m$ and $\lambda$. More precisely, we consider $m_i=m(c_i)$ and $\lambda_i=\lambda(c_i)$, $i=1,...,5$, as parameters of the model and minimize the objective function subject to the constraints
\begin{eqnarray}
&&m_1\leq m_2\leq m_3 \leq m_4 \leq m_5,\label{mon_m}\\
&&\lambda_1\leq \lambda_2\leq \lambda_3\leq \lambda_4\leq \lambda_5\label{mon_lambda}.
\end{eqnarray}

\begin{table}[ht!]
\caption{LSM fit of model \eqref{eqCelVia2} with monotonicity constraints \eqref{mon_m}-\eqref{mon_lambda} to the cell viability experimental data for B16F10, RAW264.7 and HaCaT cell populations.}\label{table1}\vspace{12pt}

\centering
\begin{tabular}{|c||c|c||c|c||c|c|}\hline &\multicolumn{2}{|c||}{}&\multicolumn{2}{|c||}{}&\multicolumn{2}{|c|}{}\\[-8pt]
&\multicolumn{2}{|c||}{B16F10}&\multicolumn{2}{|c|}{RAW264.7}&\multicolumn{2}{|c|}{HaCaT}\\[-8pt]
&\multicolumn{2}{|c||}{}&\multicolumn{2}{|c||}{}&\multicolumn{2}{|c|}{}\\\hline &&&&&&\\[-8pt]
conc.&$m$&$\lambda$&$m$&$\lambda$&$m$&$\lambda$\\[-8pt]&&&&&&\\\hline &&&&&&\\[-8pt]
1mM & $7.1491\!\!\times\!\! 10^{-4}$ & $1.4856\!\!\times\!\! 10^5$ & $ 10^{-22}$ & $3.1880\!\!\times\!\! 10^{-5}$& $5.7831\!\!\times\!\! 10^{-2}$ & $1.8830\!\!\times\!\! 10^{-3}$ \\
2mM & $1.2955\!\!\times\!\! 10^{-3}$ & $2.7832\!\!\times\!\! 10^5$& $1.9120$ & $4.0706\!\!\times\!\! 10^{-5}$& $5.7831\!\!\times\!\! 10^{-2}$ & $1.8830\!\!\times\!\!10^{-3}$\\
3mM & $2.2501\!\!\times\!\! 10^{-3}$ & $2.7832\!\!\times\!\! 10^5$& $1.9120$ & $4.0706\!\!\times\!\! 10^{-5}$& $5.7831\!\!\times\!\! 10^{-2}$ & $1.8830\!\!\times\!\! 10^{-3}$\\
4mM & $3.6713\!\!\times\!\! 10^{-3}$ & $6.6887\!\!\times\!\! 10^6$& $2.3498$ & $4.0706\!\!\times\!\! 10^{-5}$& $5.7831\!\!\times\!\! 10^{-2}$ & $3.8878\!\!\times\!\! 10^{-3}$\\
5mM & $4.0894 \!\!\times\!\! 10^{-3}$ & $1.7556\!\!\times\!\! 10^7$& $2.3498$ & $4.0706\!\!\times\!\! 10^{-5}$& $6.8093\!\!\times\!\! 10^{-2}$ & $3.8878\!\!\times\!\! 10^{-3}$ \\[-8pt]&&&&&&\\\hline
&\multicolumn{2}{|c||}{}&\multicolumn{2}{|c||}{}&\multicolumn{2}{|c|}{}\\[-8pt]
&\multicolumn{2}{|c||}{$\text{RMSE}_{\text{means}} = 4.0110$}&\multicolumn{2}{|c||}{$\text{RMSE}_{\text{means}} = 5.1023$}&\multicolumn{2}{|c|}{$\text{RMSE}_{\text{means}} = 8.4733$}\\[-8pt]
&\multicolumn{2}{|c||}{}&\multicolumn{2}{|c||}{}&\multicolumn{2}{|c|}{}\\\hline
\end{tabular}
\end{table}

The results for each of the three cell populations are given in Table \ref{table1}. These results do not yield a complete models, as we do not have $m$ and $\lambda$ as functions of $c$. However, from these initial results we can draw some very relevant conclusions.
\begin{itemize}
\item The error is a lower bound for the error in any model, where $m$ and $\lambda$ are monotone increasing. Hence, a model approximating the data with error not significantly larger than the respective error in Table \ref{table1} can be considered reasonably accurate.
\item The values and trends in Table \ref{table1} may inform the selection of a specific parametrisation of the model.
\end{itemize}

\subsubsection{The model of inhibition of B16F10.}

\begin{figure}[ht!]
\centering\begin{minipage}{6cm}
\captionof{table}{LSM fit of model \eqref{eqCelVia3} with monotonicity constraint \eqref{mon_m}.}\label{table2_B16F10}\vspace{12pt}

\centering
\begin{tabular}{|c|c|}\hline &\\[-8pt]
conc.&$m$\\[-8pt]&\\\hline &\\[-8pt]
1mM & $7.1491\times 10^{-4}$  \\
2mM & $1.2955\times 10^{-3}$ \\
3mM & $2.2501\times 10^{-3}$ \\
4mM & $3.6713\times 10^{-3}$ \\
5mM & $4.0894 \times 10^{-3}$  \\[-8pt]&\\\hline
\multicolumn{2}{|c|}{}\\[-8pt]
\multicolumn{2}{|c|}{$\text{RMSE}_{\text{means}} = 4.0110$}\\[-8pt]\multicolumn{2}{|c|}{}\\\hline
\end{tabular}
\end{minipage}\hspace{5mm}
\begin{minipage}{9cm}
\centering\includegraphics[scale=0.36]{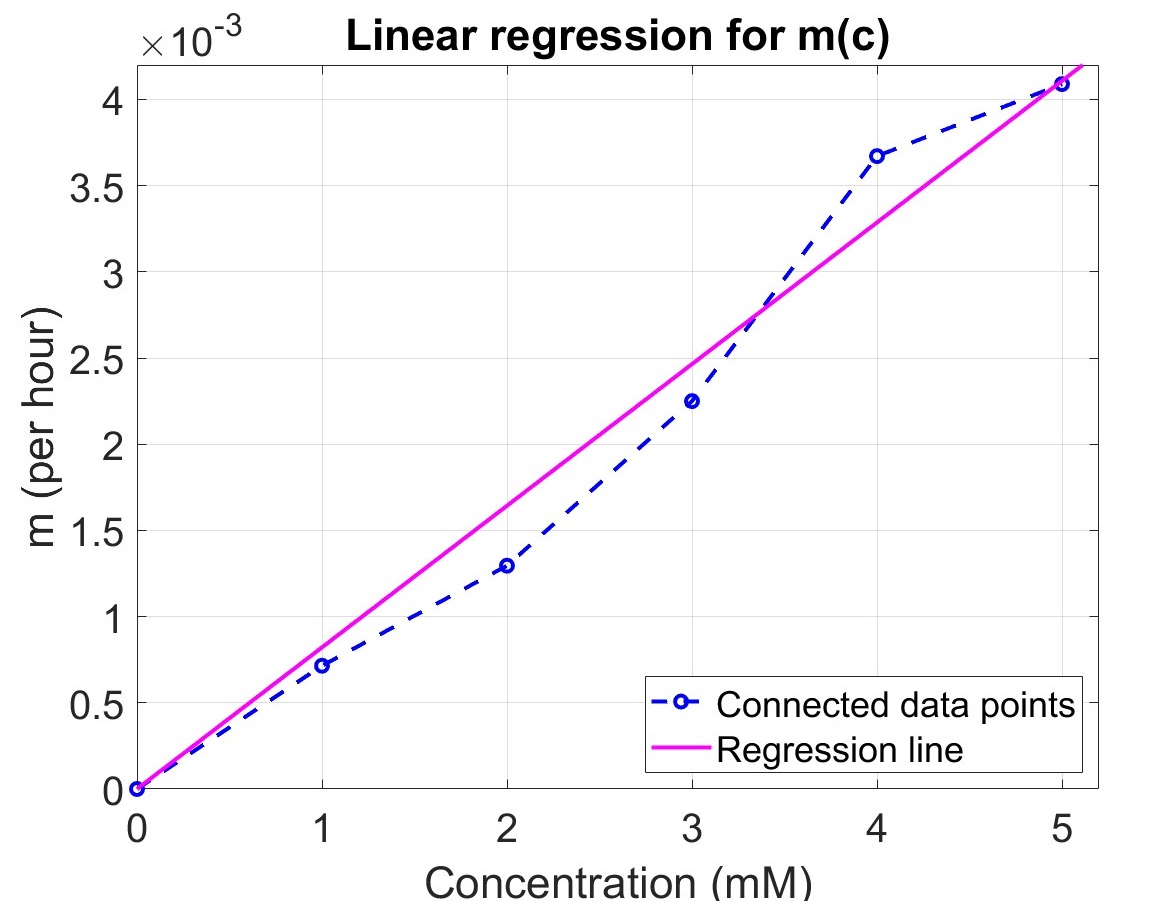}
\caption{The values of $m$ in Table \ref{table2_B16F10} and the linear regression line. Explained variation: $R^2=97.6\%$}\label{M_Regression}
\end{minipage}
\end{figure}

We observe that the values of $\lambda$ in Table \ref{table1} are very large. Hence, the function \eqref{eqCelVia2} is practically independent of $\lambda$ and the model can be reduced to \eqref{eqCelVia4}. The repeat of the fitting procedure for model \eqref{eqCelVia4} with the constraint \eqref{mon_m} yields no measurable difference of error RMSE$_{\text{means}}$ or the values of $m$, see Table \ref{table2_B16F10}. On the numerical implementation, we can remark that the minimization of the LSM objective function for the model \eqref{eqCelVia4} is robust in the sense that there is a unique local minimum, which is also the global minimum. This is not the case for the model \eqref{eqCelVia2}, where we encounter several local equilibria, some with very similar values of the objective function.

The next step is to derive an explicit form of $m$ as a function of $c$. The computed values of $m$ strongly suggest a linear relationship. This can be seen on Figure \ref{M_Regression}, where the values of $m$ as well as the regression line through the origin are presented. In this way, we obtain the following model for the cell viability function
\begin{equation}\label{eqCelVia51}
\Psi(c,t)=\varphi(a;c,t):=100e^{-act}
\end{equation}
In the linear regression represented in Figure \ref{M_Regression}, we obtain an estimated value of $a$. Since this value is a result of a two step process, to avoid accumulation of error, we fit the model \eqref{eqCelVia51} directly to the original data. The results of the fitting are as follows
\begin{equation}\label{QuinSurfaceFit}
\hat{a}=8.1663\!\times\!10^{-4},\ \   \text{RMSE}_{\text{means}}=4.1688, \ \ \text{RMSPE}=8.1709, \ \ \frac{\text{RMSPE}}{s_{\text{within}}}=1.0240.
\end{equation}
We can observe that RMSE$_\text{means}$ is in the same range as the RMSE$_\text{means}$ for B16F10 in Table \ref{table1} or Table \ref{table2_B16F10}. This is remarkable because the error in Table \ref{table1} results from fitting of a 10 parameter model to 15 data points, while the results in \eqref{QuinSurfaceFit} are obtain by fitting a one parameter model to the same data points. It is an indication that the linear model chosen for $m(c)$ is appropriate.

\subsubsection{The model of inhibition of RAW264.7.}

\begin{figure}[ht!]
\centering\begin{minipage}{6cm}
\captionof{table}{LSM fit of model \eqref{eqCelVia3} with monotonicity constraint on $\tilde{m}$.}\label{table2_RAW}\vspace{12pt}
\begin{tabular}{|c|c|}\hline &\\[-8pt]
conc.&$\tilde{m}$\\[-8pt]&\\\hline &\\[-8pt]
1mM & $4.2358\times 10^{-9}$  \\
2mM & $3.8872\times 10^{-5}$ \\
3mM & $3.8893\times 10^{-5}$ \\
4mM & $4.7770\times 10^{-5}$ \\
5mM & $4.7808 \times 10^{-5}$  \\[-8pt]&\\\hline
\multicolumn{2}{|c|}{}\\[-8pt]
\multicolumn{2}{|c|}{$\text{RMSE}_{\text{means}} = 5.1023$}\\[-8pt]\multicolumn{2}{|c|}{}\\\hline
\end{tabular}
\end{minipage}\hspace{5mm}
\begin{minipage}{9cm}
\centering\includegraphics[scale=0.36]{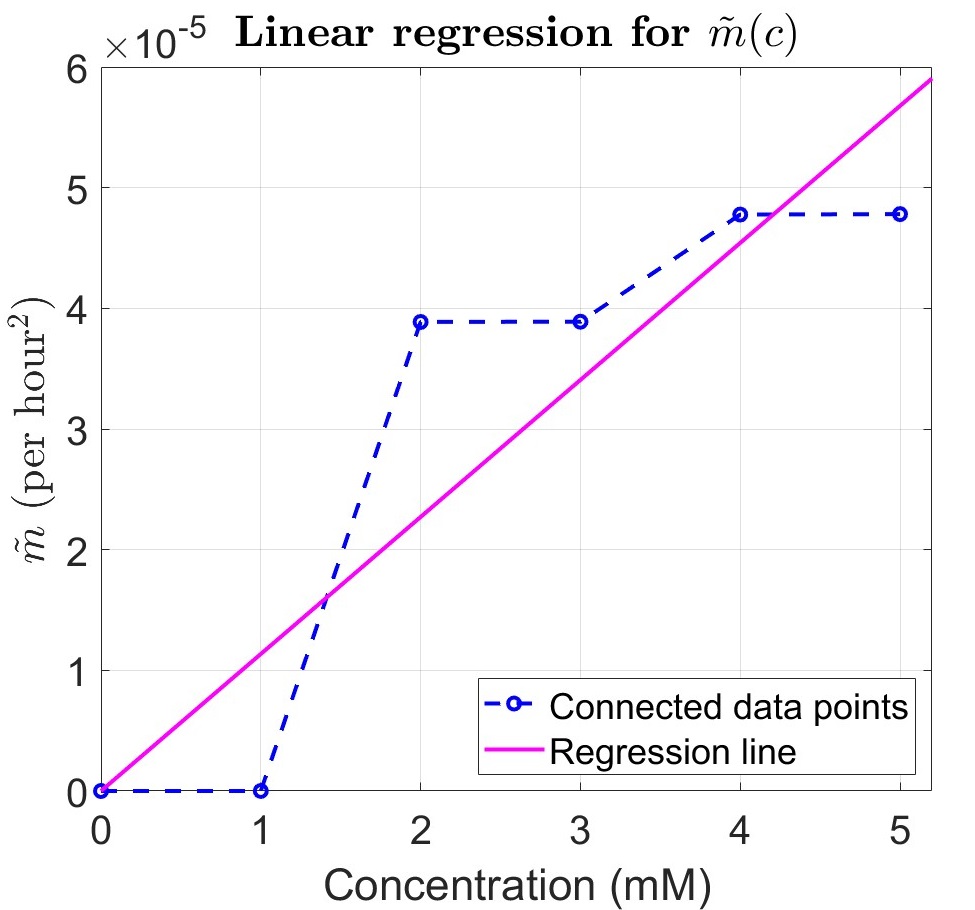}
\caption{The values of $\tilde{m}$ in Table \ref{table2_RAW} and the linear regression line. Explained variation: $R^2=80.7\%$.}\label{tildeM_Regression}
\end{minipage}
\end{figure}

We observe that the values of $\lambda$ in Table \ref{table1} are very small. Hence, the model \eqref{eqCelVia2} can be reduced to \eqref{eqCelVia3}. The repeat of the fitting procedure for model \eqref{eqCelVia3} with a monotonicity constraint on $\tilde{m}$ yields the results in Table \ref{table2_RAW}. There is no measurable change of the error of approximation.

The values of $\tilde{m}$ are increasing with $c$. The increments are rather irregular, but in the absence of other information we use the simplest approach, namely, a linear model $\tilde{m}=ac$, see Figure \ref{tildeM_Regression}. While not perfect, the linear model explains 80.7\% of the variation of $\tilde{m}$. In this way, we obtain the following model for the cell viability function
\begin{equation}\label{eqCelVia6}
\Psi(c,t)=\varphi(a;c,t):=100e^{-act^2}
\end{equation}
The results of fitting this model to the data are as follows
\begin{equation}\label{RawFinalFit}
\hat{a}=1.1205\!\times\! 10^{-5},\ \   \text{RMSE}_{\text{means}}=6.2571,\ \ \text{RMSPE}=9.8109, \ \ \frac{\text{RMSPE}}{s_{\text{within}}}=1.1694.
\end{equation}
We observe that increase of the error $\text{RMSE}_{\text{means}}$ compared to the error in Table \ref{table1} or Table \ref{table2_RAW} is relatively small, which indicates that the model \eqref{eqCelVia6} with parameter value as given in \eqref{RawFinalFit} adequately captures the data dynamics of the data.

\subsubsection{The model of inhibition of HaCaT.}

The values of $\lambda$ neither too small to reduce the model to \eqref{eqCelVia3} or too large to reduce the model \eqref{eqCelVia4}. So $\lambda$ needs to be taken into account. The parameter $m$ is assumed to be an increasing function of $c$ with $m(0)=0$. This is not really observed in the values in Table \ref{table1}. However, considering that data for the HaCaT inhibition displays high variability. Hence, it is possible that these values capture a significant amount of noise rather than trend. We accommodate all these consideration my considering $\lambda$ a constant, while $m$ increases linearly with $c$, namely $m=bc$. In this way, we obtain the following model for the cell viability function
\begin{equation}\label{eqCelVia7}
\Psi(c,t)=\varphi(a;c,t):=100e^{-bc\frac{e^{-\lambda t}-1+\lambda t}{\lambda}},
\end{equation}
where $a=(b,\lambda)$ is a two dimensional parameter.
The results of fitting this model to the data are as follows
\begin{equation}\label{HaCatFinalFit}
\hat{a}=\begin{pmatrix}3.3026\times 10^{-3}\\2.2117\times 10^{-2}\end{pmatrix},\ \   \text{RMSE}_{\text{means}}=10.3777,\ \ \text{RMSPE}=20.45, \ \ \frac{\text{RMSPE}}{s_{\text{within}}}=1.1651.
\end{equation}
The error $\text{RMSE}_{\text{means}}$ increases compared to its value in Table \ref{table1}, but not in a significant way. This indicates that the model \eqref{eqCelVia7} with parameter values given in \eqref{HaCatFinalFit}, while capturing the general dynamics of the conceptual model, \eqref{eqCelVia2} is also close to the experimental data.

\subsection{Validation of the specific inhibition models.}

\begin{figure}[ht!]
\centering\includegraphics[scale=0.32]{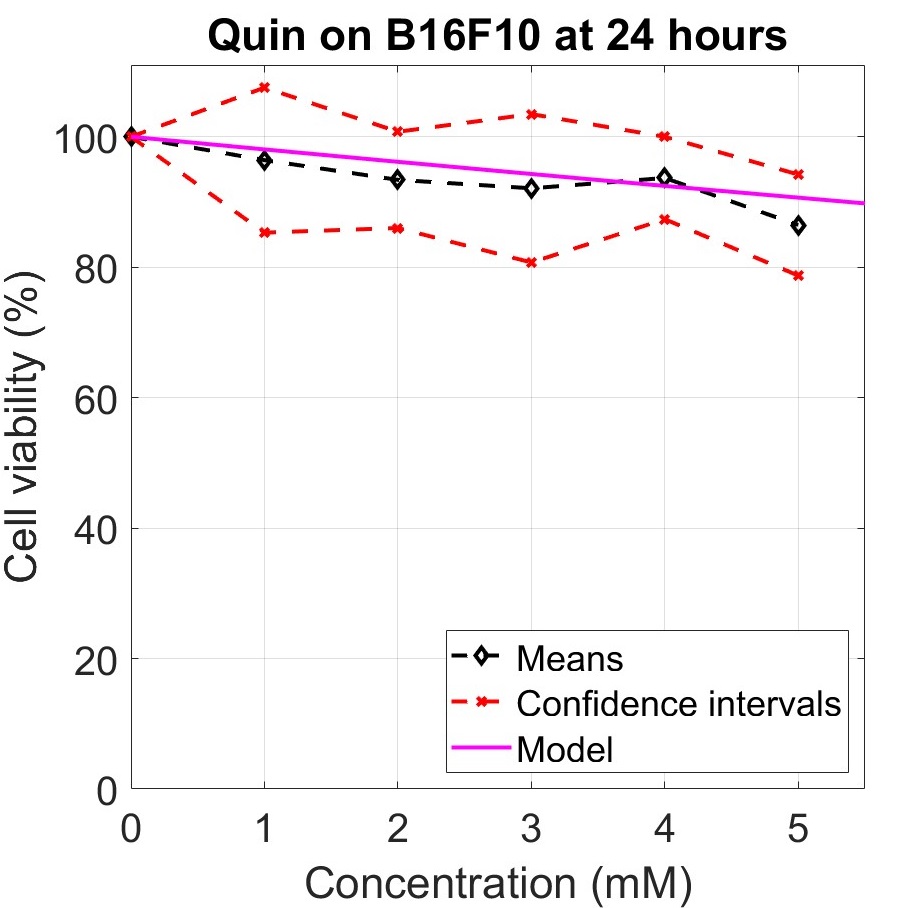} \includegraphics[scale=0.32]{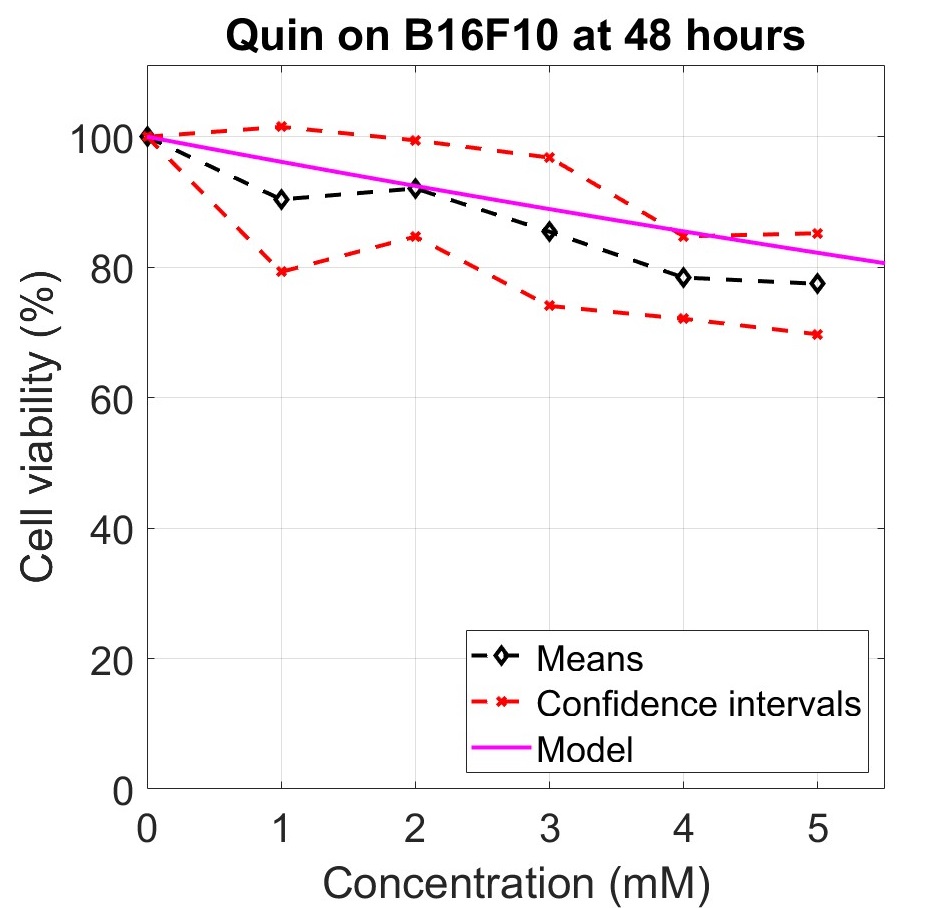}
\includegraphics[scale=0.32]{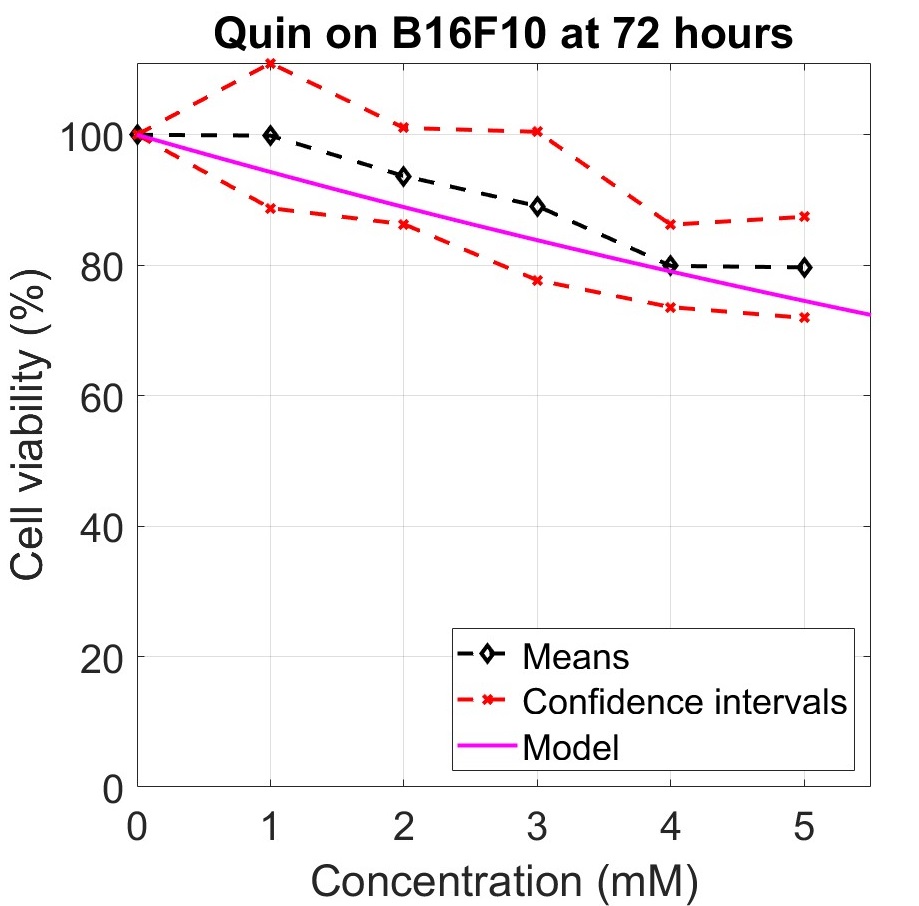}\vspace{2mm}

\centering\includegraphics[scale=0.32]{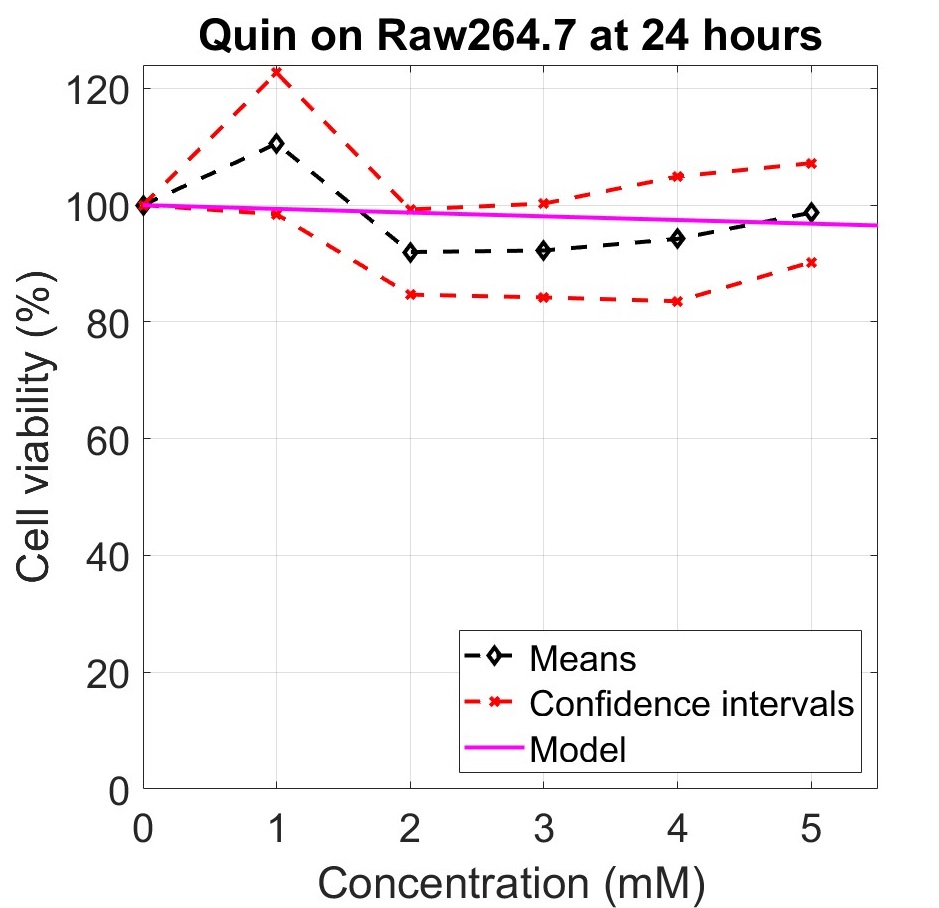} \includegraphics[scale=0.32]{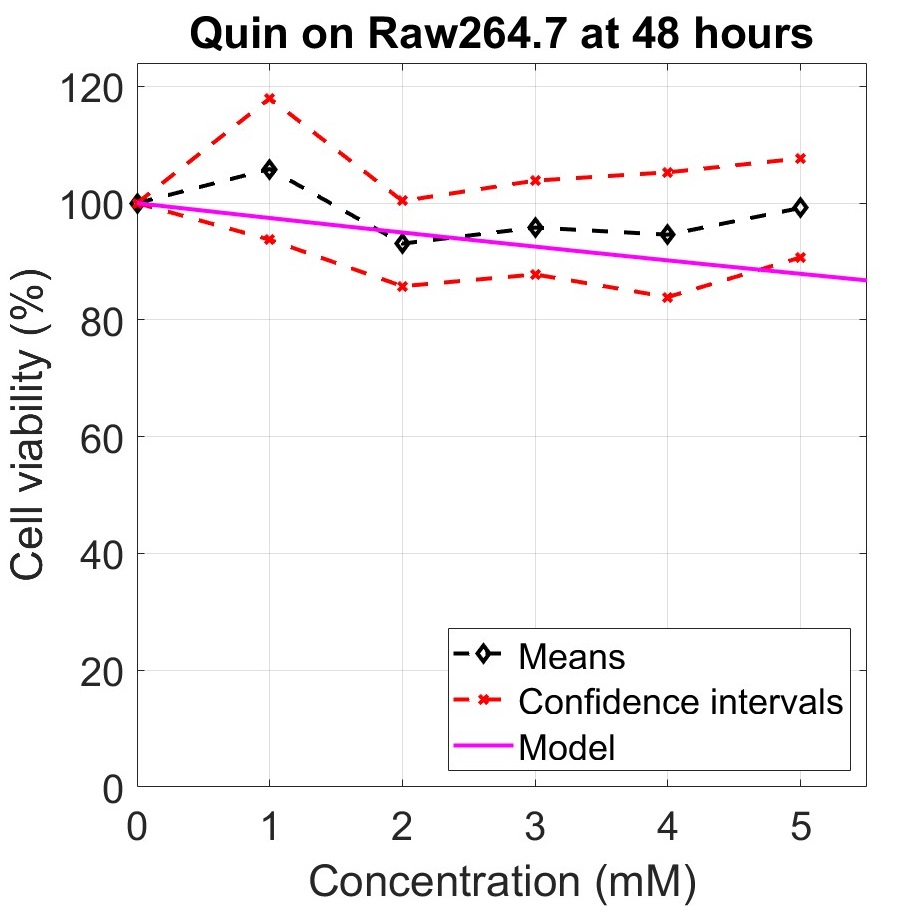}
\includegraphics[scale=0.32]{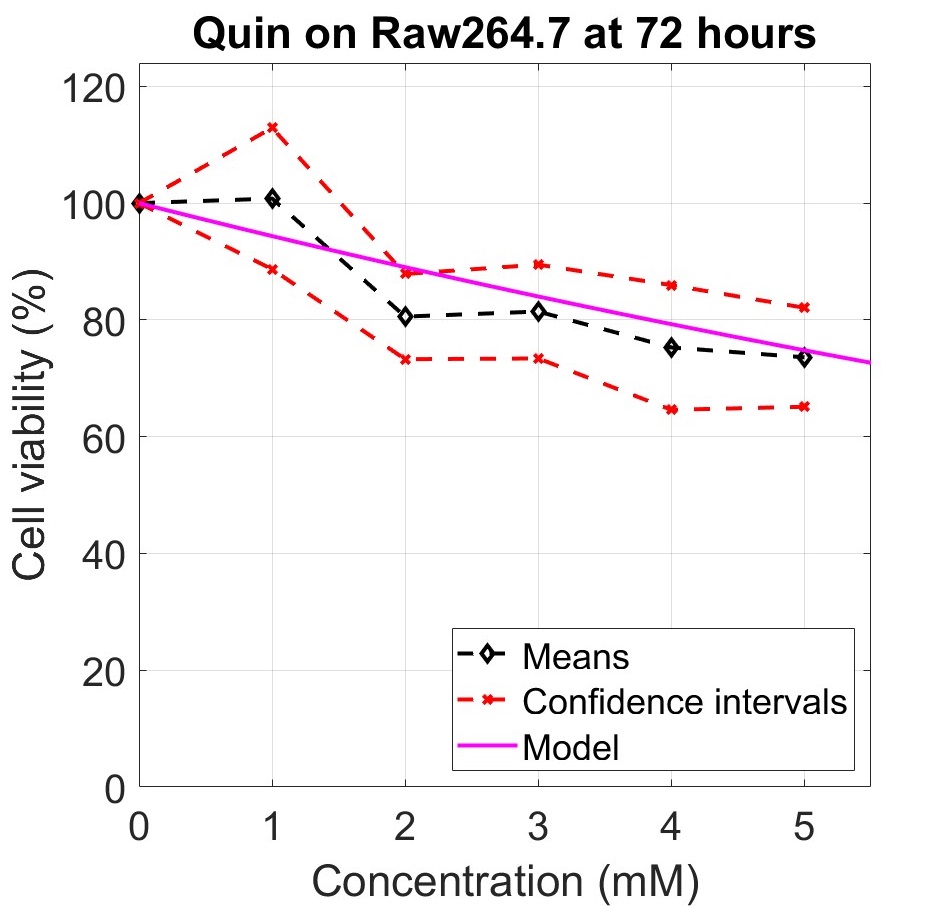}\vspace{2mm}

\centering\includegraphics[scale=0.32]{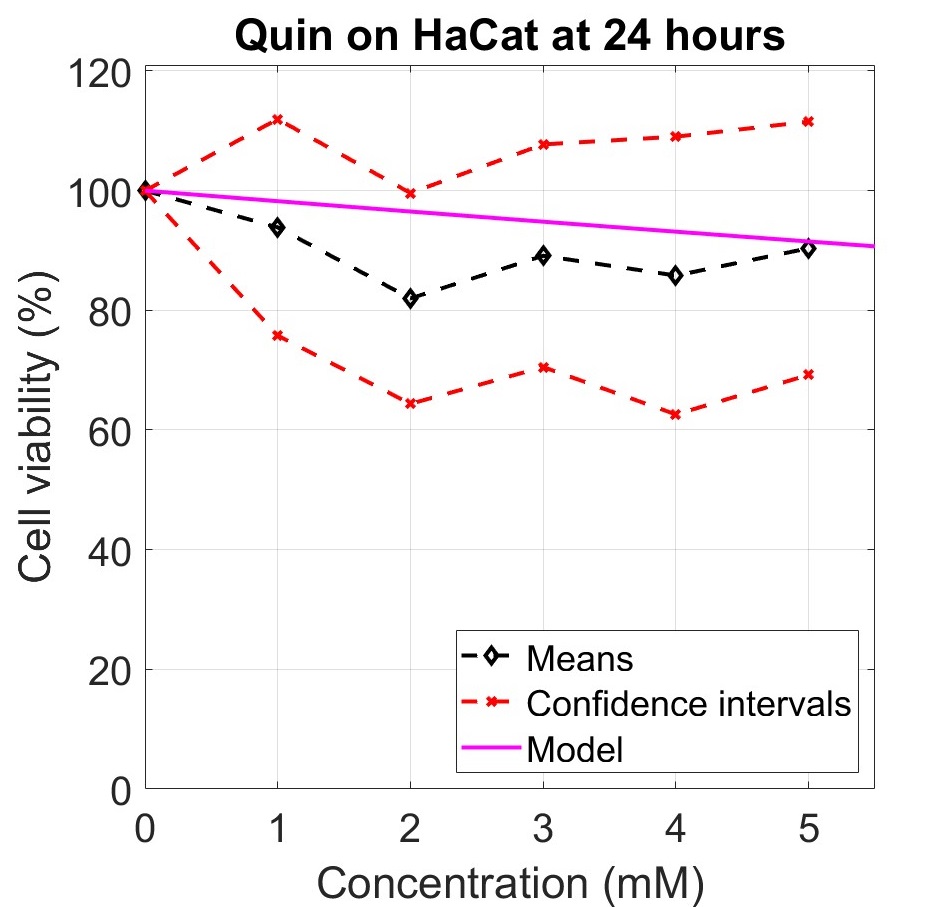} \includegraphics[scale=0.32]{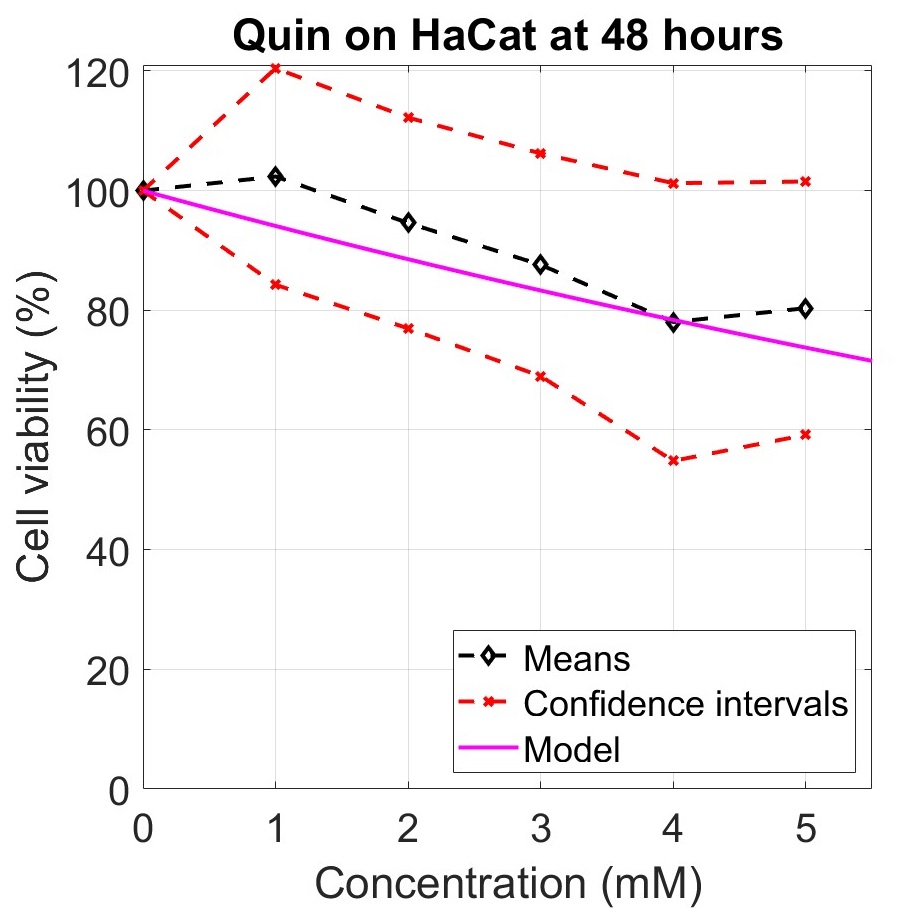}
\includegraphics[scale=0.32]{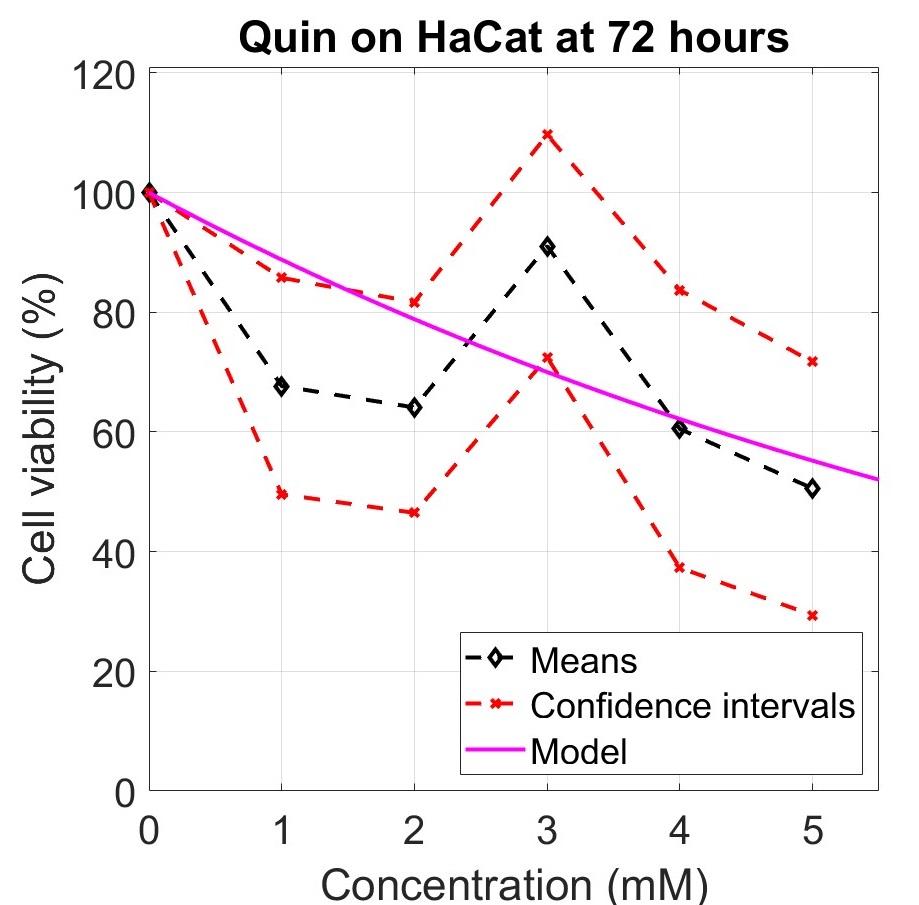}

\caption{Plots of the means, the confidence intervals and the derived model of the cell viability of cell populations.}\label{fitplots_concentration}\vspace{1mm}

\captionof{table}{Summary of models, errors, and predictive reliability for the inhibition of B16F10, RAW264.7 and HaCaT populations}\label{summary}

\centering\begin{tabular}{|c|c|c|c|c|c|}\hline &&&&&\\[-8pt]
&Model&parameter values&RMSE$_{\text{means}}$&RMSPE&$\displaystyle\frac{\text{RMSPE}}{s_{within}}$\\[-8pt]&&&&&\\\hline &&&&&\\[-8pt]
B16F10 &$100e^{-act}$ &$\hat{a}=8.1663\!\times\!10^{-4}$&4.1688&8.1709&1.0240 \\[-8pt]&&&&&\\\hline &&&&&\\[-8pt]
RAW264.7&$100e^{-act^2}$&$\hat{a}=1.1205\!\times\! 10^{-5}$&6.2571&9.8109&1.1694\\[-8pt]&&&&&\\\hline &&&&&\\[-8pt]
HaCaT&$\displaystyle 100e^{-bc\frac{e^{-\lambda t}-1+\lambda t}{\lambda}}$ &$\begin{matrix}\hat{b}=3.3026\!\times \! 10^{-3}\\\hat{\lambda}=2.2117\!\times\! 10^{-2}\end{matrix}$&10.3777&20.45&1.1651\\[-8pt]&&&&&\\\hline
\end{tabular}
\end{figure}

We recall that the 95\%-confidence intervals in Section \ref{Subsection_CI} were derived independently of any model.
Figure \ref{fitplots_concentration} represents the graphs of the models superimposed on the plots of the confidence intervals and means.  We observe that the graph of the models mostly are within the 95\%-confidence interval for the mean of the cell viability. Deriving a precise statistical measure is problematic due to various dependencies. Nevertheless, the graphs in Figure \ref{fitplots_concentration} visualise a good fit of the models to the experimental data.

The results of the model fitting are summarised in Table \ref{summary}. Considering that the units of the error RMSE$_\text{means}$ are the same as the units of the cell viability function, namely percentage of the control population, the relatively small values of RMSE$_\text{means}$ indicate a good accuracy of the model. Further, as discussed in Section \ref{sec_validation}, the size of RMSPE compared to $s_{\text{within}}$ is a measure of predictive reliability. We observe in Table \ref{summary} that for all three models the ratio $\displaystyle\frac{\text{RMSPE}}{s_{within}}$ is consistently very close to one. The values of the other indicators RMSE$_\text{means}$, RMSPE vary more substantially as a result of the change in the variability of the data across cell population, e.g observe the change of width of the confidence intervals across cell population, while the ratio $\displaystyle\frac{\text{RMSPE}}{s_{within}}$ is in the interval $[1,1.17]$ for all cell populations. This is a good indicator that the specific models in all three settings capture reliably the trend of the respective experimental data.

\section{Discussion}

All models for the three settings considered here are all derived from the general model \ref{eqCelVia2}. In turn, the validity of these models presented in the preceding section confirms that the general model \ref{eqCelVia2} provides a good guidance to quantifying the dynamics of the inhibition of cancer cells. The process of model derivation showed that it is rather important to have such guiding model, when we are have variable experimental data. The specific models as given in Table \ref{summary} are different, but they are derived from \ref{eqCelVia2} and parameterised using as few as possible parameters.

The practical applications of the models is to provide reliable predictions, which reduce the amount of the required experimental work. The predictive reliability of the model is demonstrated in Section \ref{sec_validation}. This means that, given concentration of the inhibitor $c$ and time $t$, the model provides a reliable prediction for the cell viability $\Psi(c,t)$.  Since the models provide an explicit relationship between concentration $c$, time $t$ and cell viability (CV), given any of the two of these quantities one can calculate the third one. The most common question is what concentration yields 50\% cell viability at given time. This concentration is called $IC_{50}(t)$, or if the time is already specified, just $IC_{50}$. In more general form, the question is: given cell viability CV and time $t$, find concentration $c$ such that $\text{CV}=\Psi(c,t)$. Explicit formulas for $c$ in terms $t$ and CV are given in Table \ref{table_conc}.

\begin{table}[h!]
\caption{The concentration made subject of the model equations given in Table \ref{summary}: $c$ function of CV and $t$, and specifically expressions for $IC_{50}(t)$.}\label{table_conc}
\centering\begin{tabular}{|c|c|c|}\hline &&\\[-8pt]
&Concentration as function of CV and $t$&$IC_{50}$\\[-8pt]&&\\\hline &&\\[-8pt]
B16F10&$\displaystyle c=-\frac{1}{\hat{a}t}\ln \frac{CV}{100}$&$\displaystyle IC_{50}(t)=\frac{\ln 2}{\hat{a}t}$\\&&\\
RAW264.7&$\displaystyle c=-\frac{1}{\hat{a}t^2}\ln \frac{CV}{100}$&$\displaystyle IC_{50}(t)=\frac{\ln 2}{\hat{a}t^2}$\\&&\\
HaCaT&$\displaystyle c=-\frac{\hat{\lambda}}{\hat{b}(e^{-\hat{\lambda} t}-1+\hat{\lambda} t)}\ln\frac{CV}{100}$&$\displaystyle IC_{50}(t)=\frac{\hat{\lambda}\ln 2}{\hat{b}(e^{-\hat{\lambda} t}-1+\hat{\lambda} t)}$\\[-8pt]&&\\\hline
\end{tabular}
\end{table}

\begin{figure}[ht!]
\begin{minipage}{9cm}
\includegraphics[scale=0.4]{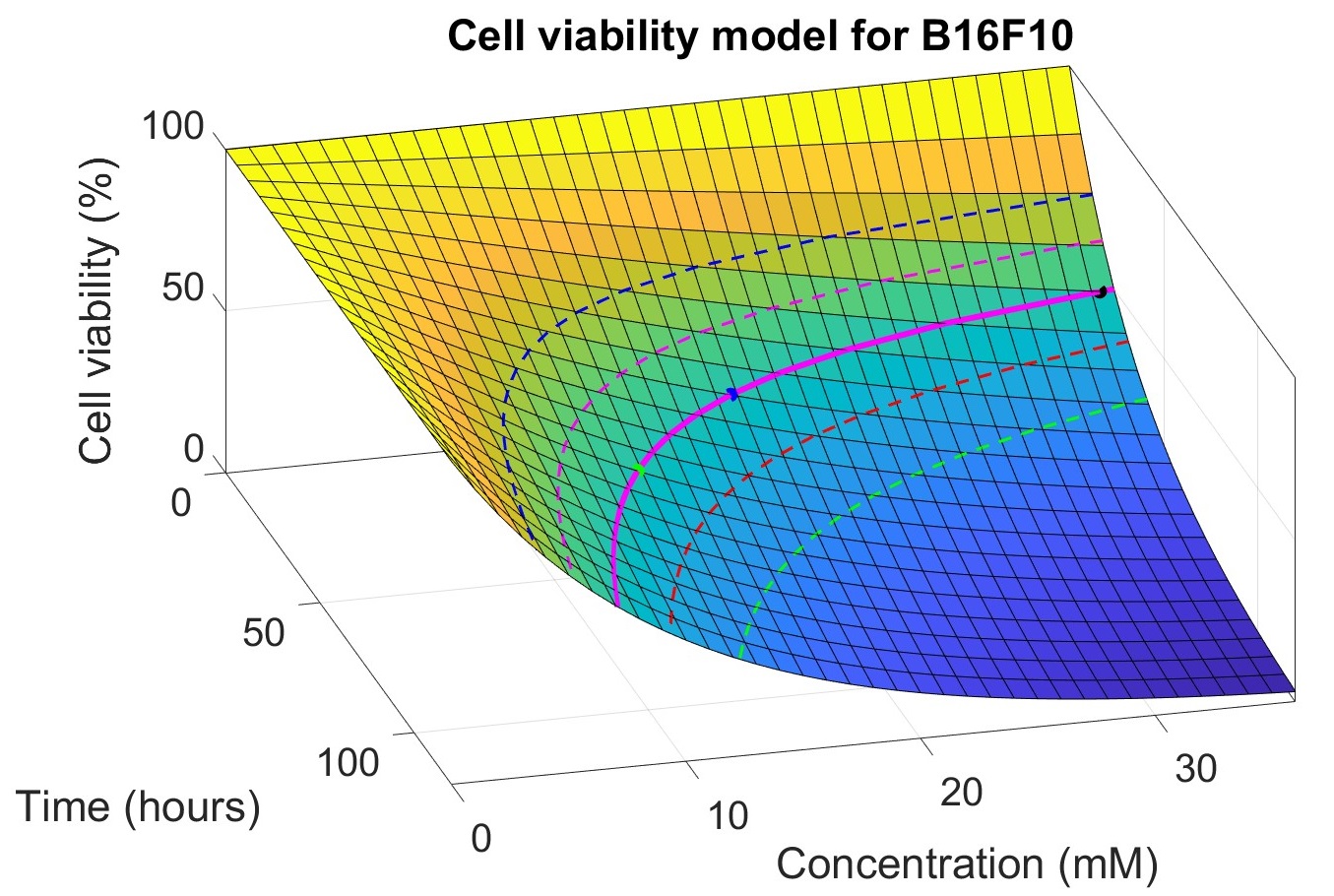}
\label{surf_B16F10}
\end{minipage}\ \ \ \ \ \ \ \ \
\begin{minipage}{6.5cm}
\includegraphics[scale=0.37]{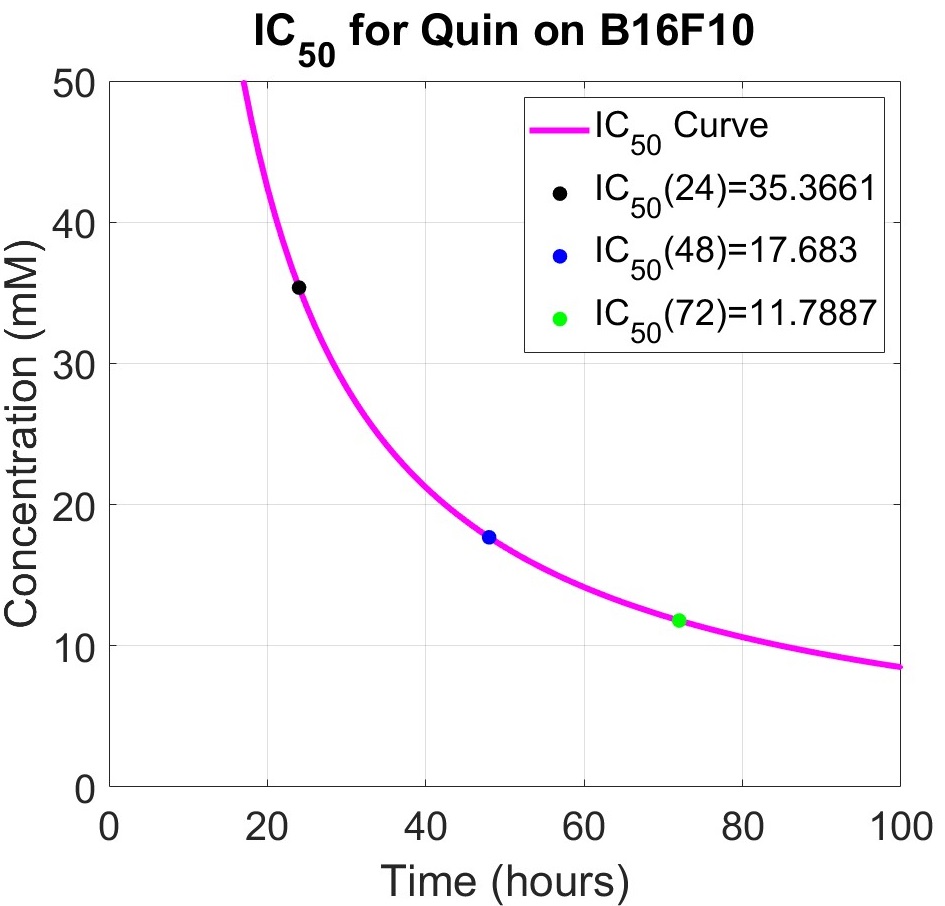}
\end{minipage}

\begin{minipage}{9cm}
\includegraphics[scale=0.4]{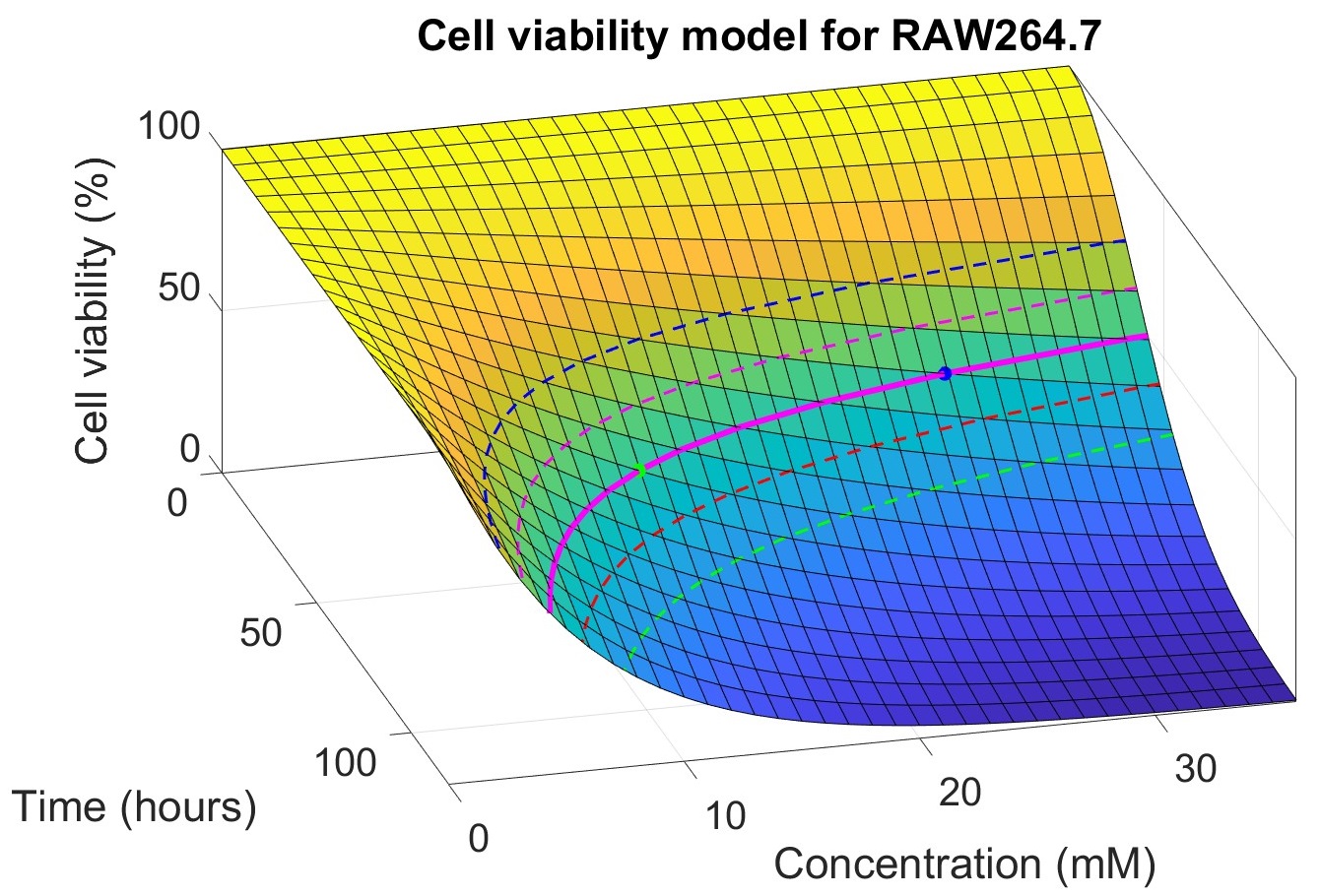}
\end{minipage}\ \ \ \ \ \ \ \ \
\begin{minipage}{6.5cm}
\includegraphics[scale=0.35]{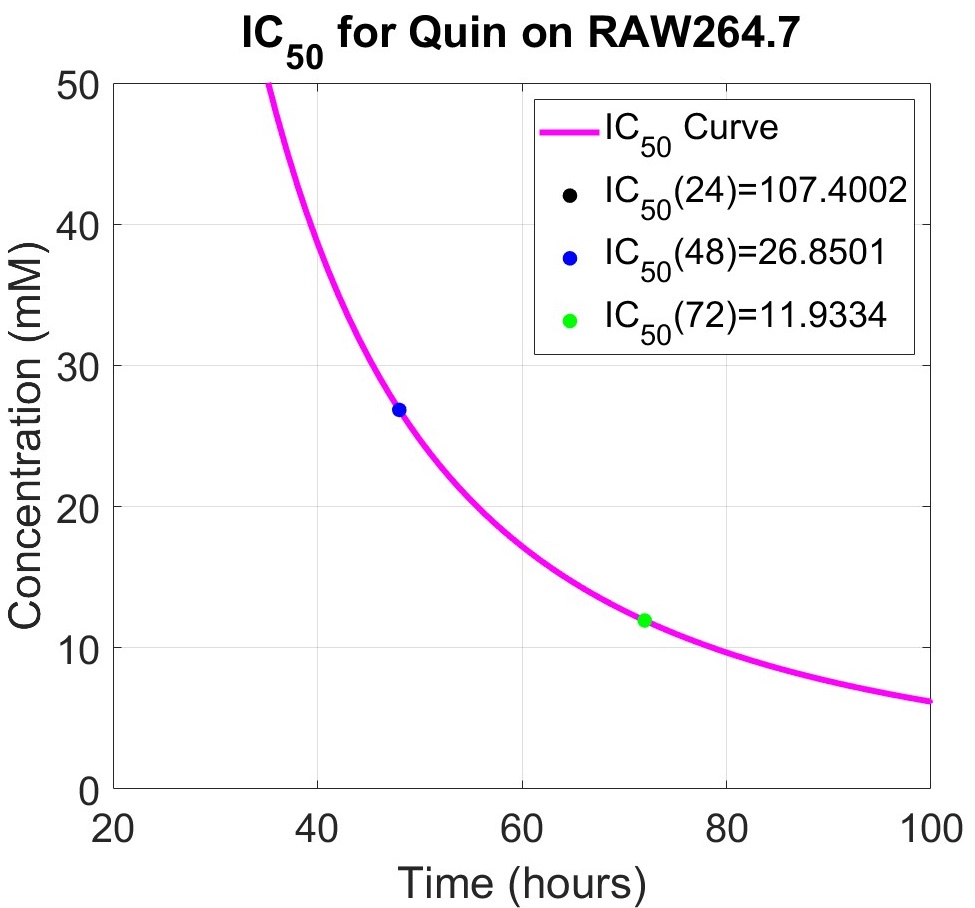}
\end{minipage}

\begin{minipage}{9cm}
\includegraphics[scale=0.4]{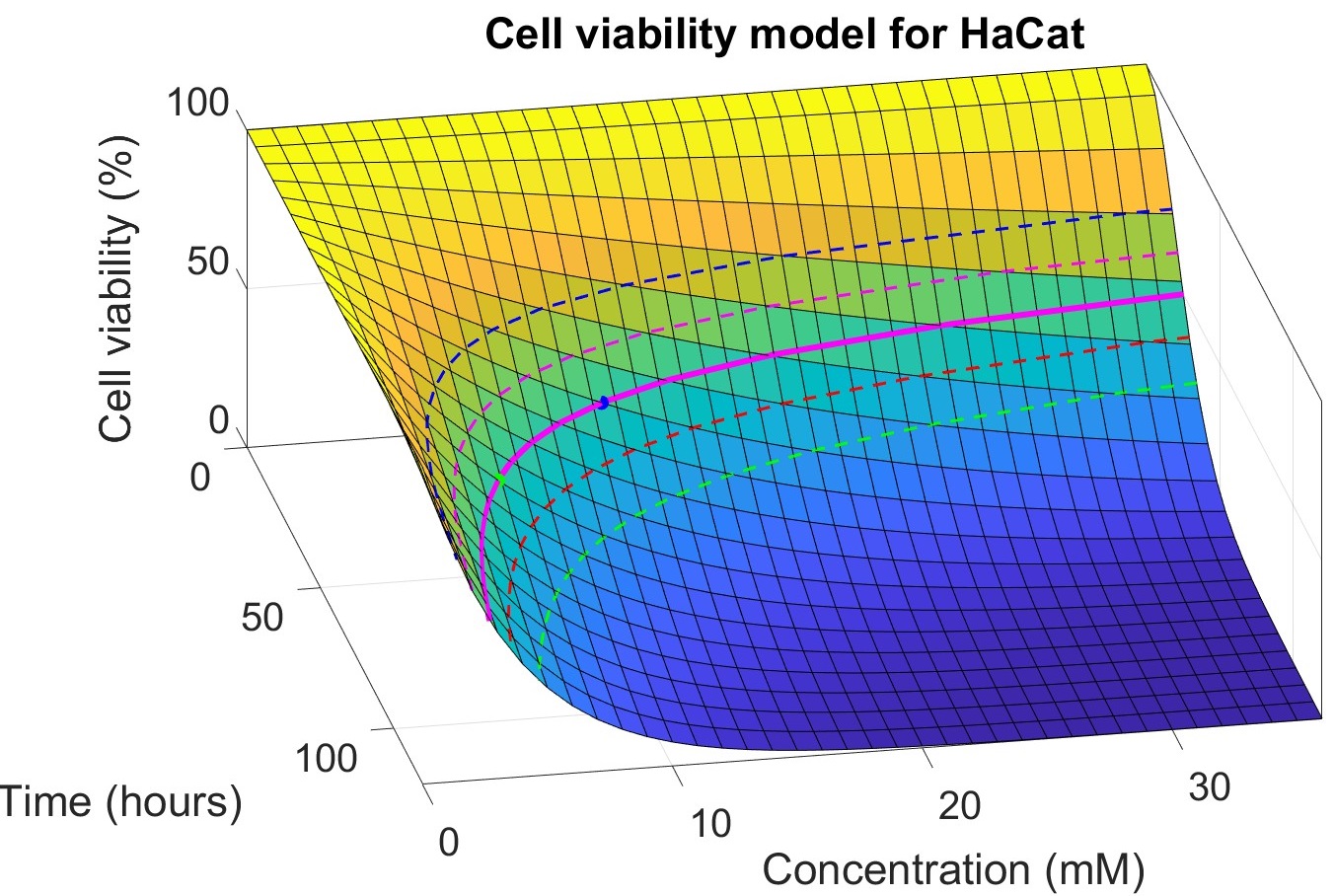}
\caption{Surface plot of the cell viability vs. concentration and time. Level lines at cell viability of 30\%, 40\%, 50\%, 60\%, 70\% determine $IC_{30}$, $IC_{40}$, $IC_{50}$, $IC_{60}$, $IC_{70}$ as functions of time.}\label{surf_Quin}
\end{minipage}\ \ \ \ \ \ \ \ \
\begin{minipage}{6.5cm}
\includegraphics[scale=0.35]{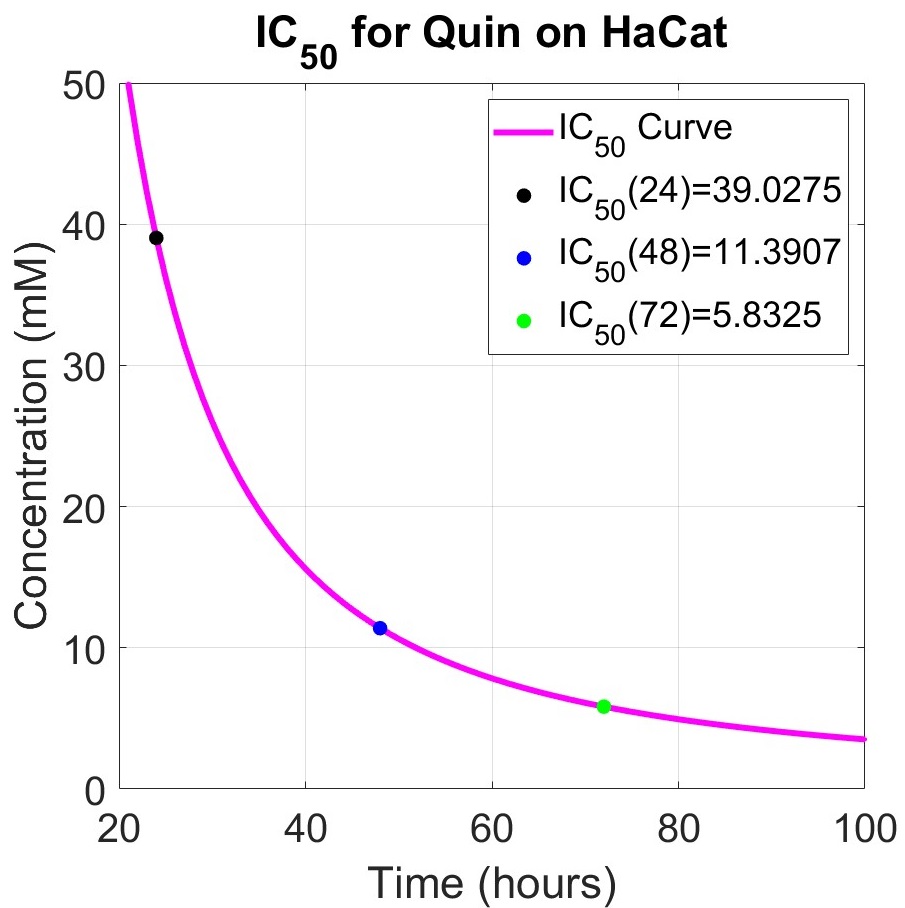}
\caption{The concentration $IC_{50}$ as a function of $t$. The values of $IC_{50}(24)$, $IC_{50}(48)$ and $IC_{50}(72)$ are specifically calculated and plotted. }\label{surf_IC50}
\end{minipage}
\end{figure}

Surface graphs of the three models of $\varphi(\hat{a};c,t)$ presented in Figure \ref{surf_Quin} visualize the properties of the models.
the level curves at fixed CV define the relationship between time and concentration. The level curves at 30\%, 40\%, 50\% 60\% and 70\% are plotted on the surfaces in Figure \ref{surf_Quin}. They visually represent the relationship between $c$ and $t$ as given Table \ref{table_conc} for CV fixed at the specified level.  The plots in Figure \ref{surf_IC50} specifically provide for each model $IC_{50}$ as a function of time  as well as its values at specific times, which are usually of interest.

\section{Conclusion}

In this paper we have developed and validated mathematical models for the antiproliferative effect of Quin on three cell lines: B16F10 melanoma, RAW264.7 macrophages, and HaCaT keratinocytes. The experimental data exhibited substantial variability and violated the independence assumption required for classical inferential statistics. To overcome these challenges, we adopted a hybrid methodology that separates the problem into a minimal statistical component (construction of model-free confidence intervals and calculation of the within-replicate standard deviation) and a deterministic approximation component (least-squares fitting to the experimental means and leave-one-replicate-out cross-validation).

Despite using the same general model (exp-integral form), the three cell lines require different parametrizations (exponential in $ct$, exponential in $ct$², and a two-parameter model). This reflects differences in their sensitivity and the rate at which inhibition saturates. The models are simple (one or two parameters) yet capture the inhibition dynamics well, as shown by the $\frac{\text{RMSPE}}{s_\text{within}}$ ratios close to 1. The high variability in HaCaT ($s_\text{within} = 17.55$) suggests that this cell line is more prone to experimental fluctuation, yet the model still achieves acceptable predictive reliability. The explicit formulas for IC50 as a function of time are valuable for dose-response predictions.

The main strength of our approach is its robustness to violations of classical statistical assumptions, making it suitable for small and noisy data sets typical of preclinical drug screening. The deterministic validation via LOOCV offers a transparent and assumption-free assessment of predictive reliability. A limitation is the reliance on a single cross-validation split (with only four replicates) — while this is deterministic, the estimate of prediction error has no associated confidence interval. However, given the small number of replicates, more sophisticated re-sampling methods are not feasible. Future work could extend the framework to include model selection criteria that account for parameter uncertainty, or to incorporate prior biological knowledge as constraints.

In summary, the models presented here provide a quantitative and predictive framework for characterizing quinolinic acid–induced toxicity across three cell types, while the methodology offers a robust template for conducting similar analyses in other experimental systems.

\section*{Funding}
The research was supported by the DST/NRF SARChI Chair on Mathematical Models and Methods in Bioengineering and Biosciences at the University of Pretoria.

\section*{Data availability statement}
All data is graphically represented in Figure \ref{Data_B16F10}, Figure \ref{Data_RAW} and Figure \ref{Data_HaCat}. Numerical values are available from the authors on request.

\end{document}